\documentclass[aps,amsmath,amssymb,showpacs,showkeys]{revtex4-2}
\usepackage[dvips]{graphicx}
\usepackage{times}
\usepackage{braket}
\usepackage{xcolor}
\usepackage{orcidlink}
\usepackage{hyperref}
\usepackage{booktabs}
\usepackage{multirow}
\hypersetup{
  colorlinks=true,
  urlcolor=blue,
  linkcolor=red,
  citecolor=blue
}

\begin{document}
\title{Gravitational Waves from Ringing Dyonic Black Holes and Estimation of
Parameters in Einstein-Bumblebee-Maxwell Gravity}

\author{Sikha Gogoi\orcidlink{0009-0000-8433-788X}}
\email[Email: ]{sikhagogoi001@gmail.com}
\author{Umananda Dev Goswami\orcidlink{0000-0003-0012-7549}}
\email[Email: ]{umananda@dibru.ac.in}

\affiliation{Department of Physics, Dibrugarh University, Dibrugarh 786004, 
Assam, India}

\begin{abstract}
In this work, we investigate the quasinormal mode (QNM) frequencies of dyonic 
Reissner-Nordstr\"om (RN)-like black holes associated with scalar and 
gravitational perturbations in the framework of Einstein-bumblebee-Maxwell 
gravity theory. The QNMs are calculated using the 6th-order Pad\'e averaged 
WKB approximation method. We analyse the dependence of oscillation frequencies 
and damping rate of the BHs for the free model parameters. To further verify 
the QNM spectrum, we study the time evolution of both the scalar and 
gravitational perturbations around the BHs and extract the dominant QNM from 
the resulting ringdown signal. Moreover, we try to estimate the BH parameters 
using Fisher analysis method for the future gravitational wave detector named 
Laser Interferometer Space Antenna (LISA). We found that for supermassive BHs 
with mass around $10^6$ -- $10^8 M_\odot$ and at lower redshifts, the 
signal-to-noise ratio for the detection of a QNM signal by the LISA is high. 
Further, for a representative source at a luminosity distance of $3$ Gpc, the 
BH mass at the detector reference frame and the Lorentz violating parameter 
are well constrained for supermassive BHs with mass $\ge 10^6.$ 
\end{abstract}

\keywords{Gravitational waves: Quasinormal modes; Dyonic black holes; 
Einstein-bumblebee-Maxwell gravity; Estimation of black hole parameters;
Gravitational wave detector: LISA}

\maketitle                                                                      

\section{Introduction}
The first direct detection of gravitational waves (GWs) by the LIGO 
collaboration in 2015 \cite{GW_1st_detection} and subsequent detections of 
these waves from different sources \cite{GW_detection2, GW_detection3, GW_detection4, GW_detection5}, and the images of black holes (BHs) 
at the center of galaxies captured by the Event Horizon Telescope (EHT) since 
2019 \cite{BH_detection1,BH_detection2,BH_detection3,BH_detection4,
BH_detection5,BH_detection6} have been regarded as the significant achievements
in the history of physics and provide strong observational support for general 
relativity (GR) in addition to earlier supports provided to GR by various 
experimental observations. Despite all these successes, the validity of GR as a 
fundamental theory of gravity faces several challenges. For example, GR faces 
challenges in explaining several cosmological observations, including the 
missing mass problem \cite{missing_mass}, and accelerated expansion of the 
Universe \cite{acc_expansion} without invoking dark matter and dark energy, 
respectively. Moreover, GR is not renormalizable at the UV scale and hence 
provides a classical description of gravity only. On the other hand, the 
Standard Model (SM) of particle physics provides a remarkably successful 
quantum description of elementary particles and three of the four fundamental 
interactions of nature \cite{SM}. So the unification of the two theories 
describing nature at the classical and quantum levels would provide a deeper 
understanding of all the four fundamental interactions at the quantum level. 
Thus, towards this pursuit, some quantum gravity (QG) models have been 
formulated, but they require an enormous amount of energy, far beyond current 
experimental facilities, for their experimental testing. 
Consequently, direct experimental test of most of the QG theories remain 
challenging. Nevertheless, there are some effects of QG which may be observed 
at current low energy scales. One such effect is the Lorentz symmetry 
breaking, which has been considered in various approaches to QG models, 
including loop quantum gravity (LQG) \cite{LQG, LQG2}, string theory 
\cite{string_theory1, string_theory2}, 
standard model extension (SME) \cite{SME1, SME2, SME3}, etc. Among these 
approches, SME is a general field theoretical framework that includes the 
fields of GR and SM. Several studies have investigated the consequences of 
Lorentz symmetry breaking in the gravitational sector of SME (see 
e.g.~\cite{SME4, SME5, SME6}). The possible effects of Lorentz symmetry 
breaking on GWs have also been studied in Refs.~\cite{LSB1, LSB2}.\\
\indent Among these extended gravitational field theories with spontaneous 
Lorentz symmetry breaking, Einstein-bumblebee gravity is one of the simplest 
vector-tensor models \cite{SME3,bumblebee1}. In this theory, a vector field 
referred to as the bumblebee field couples with the spacetime and modifies the 
gravitational dynamics. Since its formulation, Einstein-bumblebee gravity 
theory has attracted considerable research interest. A variety of exact BH 
solutions,  including Schwarzschild-like BHs with a cosmological constant 
\cite{BH_solution1}, electrically charged RN-like BHs \cite{BH_solution2}, 
Taub-Newman-Unti-Tamburino (Taub-NUT)-like BHs \cite{BH_solution3}, etc.~have 
been subsequently investigated within Einstein-bumblebee gravity theory in 
terms of various properties of BHs and GWs \cite{bumblebee2, bumblebee3, 
bumblebee4, bumblebee5, bumblebee6, bumblebee7, bumblebee8, DJ}. In particular,
the effects of the Lorentz-violating parameter on the BH perturbations and GW 
signals have attracted significant interest, as they may provide a useful 
framework to probe deviation from GR through GW observations.\\
\indent Supermassive BHs (SMBHs) are leading sources of detectable GWs. The 
GWs associated with a binary BH merger event comprise three phases, namely,
an inspiral waveform, a merger waveform and a ringdown waveform. The inspiral 
waveform originates from the gradual orbital evolution of the binary and can 
be analyzed using post-Newtonian theory \cite{inspiral1, inspiral2}, while the 
merger waveform is highly unknown at present. Following the merger, the 
distorted final BH emits ringdown GWs as it relaxes towards a stationary state. 
The ringdown GW is characterized by the superposition of damped oscillations 
known as quasinormal modes (QNMs). QNMs are complex frequencies associated 
with perturbed compact and massive objects \cite{QNM1}.  The real part of a 
QNM frequency determines the oscillation frequency, whereas its imaginary part 
characterizes the damping rate of the perturbation. Consequently, any 
modification of the BH metric induced by the underlying theory of gravity can 
leave distinct imprints on the QNM spectrum. Since these QNMs are directly 
related to the ringdown phase of GWs, they provide a promising probe to test 
a theory through GW signals. In recent years, the properties of GWs and BH 
QNMs have been extensively investigated within various modified theories of 
gravity \cite{QNM2, QNM3, QNM4, QNM5,QNM6, QNM7, QNM8}. In particular, the 
effect of Lorentz-violating parameter on QNM frequencies of BHs in the 
framework of bumblebee gravity has also been studied in 
Refs.~\cite{bumblebee4, bumblebee7, DJ}.\\
\indent In this work, we consider a class of dyonic RN-like BHs carrying both 
electric and magnetic charges derived within Einstein-bumblebee-Maxwell (EbM) 
theory introduced in Ref.~\cite{our_BH}. The EbM theory is an extension of 
Einstein-bumblebee gravity theory, which incorporates a Maxwell field into the 
gravitational sector and gives rise to electric and magnetic charges of the 
BHs. We investigate the QNMs of the perturbed BHs by considering scalar and 
gravitational perturbations, and study the effect of the model parameters on 
the QNM spectrum. To calculate QNMs, we employ 6th order Pad\'e averaged WKB 
approximation described in Refs.~\cite{DJ,QNM6, QNM7, QNM8}. Although the WKB 
approximation method provides a good approximate calculation of QNM 
frequencies, it may sometimes not give accurate results \cite{WKB}. To compare 
our WKB approximated QNMs in the time domain, we study the time evolution of 
perturbation profiles and obtain the QNMs from time-domain analysis. 
Furthermore, we construct the ringdown gravitational waveform with the QNM 
frequencies obtained from the WKB approximation and investigate the 
detectability of the GW signal using the sensitivity of Laser Interferometer 
Space Antenna (LISA) \cite{LISA1}. We then employ the Fisher analysis to 
estimate the uncertainties in the BH parameters with future LISA 
observations.\\
\indent The paper is organized as follows. In  the next Section \ref{sec2}, we 
give a brief overview of the field equations of the EbM gravity model and the 
dyonic RN-like BH solution obtained in this gravity model. In Section 
\ref{sec3}, we formulate the corresponding scalar and gravitational 
perturbation equations, calculate the QNM frequencies using the sixth-order 
Pad\'e-averaged WKB approximation method and investigate their dependence on 
the BH parameters. In Section \ref{sec4}, we study the time-domain evolution 
of the perturbations and compare the QNM frequencies extracted from the 
time-domain profiles with those obtained from the WKB approximation method. In 
Section \ref{sec5}, we study the possibility of experimental detection of 
QNMs. In Section 
\ref{sec6}, we construct the ringdown gravitational waveform and investigate 
its detectability using the LISA sensitivity curve, and employ the Fisher 
analysis technique to estimate the uncertainties in the BH parameters and 
constrain the model parameter. Finally, Section \ref{sec7} summarizes our 
main results and presents our conclusions.
\section{Einstein-bumblebee-Maxwell gravity and the black hole solutions}
\label{sec2}
In this section, we briefly discuss the formalism of Einstein-bumblebee-Maxwell
gravity theory and the BH solutions in this theory derived in 
Ref.~\cite{our_BH}. The action of this theory is given by
\begin{equation}{\label{action}}
S=\frac{1}{2\kappa}\int d^{4}x\,\sqrt{-g}\,\left(\mathcal{L}_{1}+\mathcal{L}_{2}\right),
\end{equation}
where $\kappa=8\pi G/c^4$, $\mathcal{L}_1$ is the Lagrangian of the 
Einstein-bumblebee gravity defined as \cite{SME3}
\begin{equation}\label{Lagrangian1}
\mathcal{L}_{1} = R + \gamma R_{\mu\nu} B^{\mu} B^{\nu}
- 2\kappa \left(\frac{1}{4} B_{\mu\nu} B^{\mu\nu} + V\!\left(B_{\mu} B^{\mu}\right)\right)
\end{equation}
with $B^{\mu}$ as a vector field referred to as the bumblebee field, which 
triggers spontaneous symmetry breaking and its field strength is given by
\begin{equation}
B_{\mu\nu}={\partial}_{\mu}B_\nu-{\partial}_\nu B_\mu.
\end{equation}
The parameter $\gamma$ in Eq.~\eqref{Lagrangian1} is the coupling constant 
associated with the non-minimal interaction between the bumblebee field and the Ricci tensor (representing the gravity field), and $V$ is the potential 
associated with the bumblebee field, which can generally be expressed as
\begin{equation}
V \equiv V\!\left(B_{\mu}B^{\mu} \pm b^{2}\right)
\end{equation}
with $b^2$ as a real positive constant. Similarly, $\mathcal{L}_2$ is the 
Lagrangian associated with the electromagnetic sector governed by Maxwell 
theory, and the electromagnetic field also exhibits a non-minimal coupling 
to the bumblebee field \cite{BH_solution2, Maxwell1, Maxwell2}, as given by
\begin{equation}\label{Lagrangian2}
\mathcal{L}_{2} = -\frac{1}{4}F_{\mu\nu}F^{\mu\nu}
+\gamma_{1}B_{\lambda}B^{\lambda}F_{\mu\nu}F^{\mu\nu}
+\gamma_{2}B^{\mu}F_{\mu\nu}B_{\lambda}F^{\lambda\nu}.
\end{equation}
Here, $F_{\mu\nu}$ is the field strength tensor of the electromagnetic field 
$A_\mu$, given by
\begin{equation}
F_{\mu\nu}=\partial_\mu A_\nu-\partial_\nu A_\mu,
\end{equation}
and $\gamma_1$, $\gamma_2$ are coupling constants that characterize the 
non-minimal interaction between the electromagnetic field and the bumblebee 
field. The variation of the action \eqref{action} with respect to metric tensor 
$g_{\mu\nu}$, bumblebee field $B_\nu$ and the electromagnetic field $A_\nu$, 
gives respectively the following field equations \cite{our_BH}:
\begin{align}
E_{\mu\nu} \equiv {} & G_{\mu\nu}+\gamma \Bigg[-\frac{1}{2} g_{\mu\nu} R_{\rho\sigma} B^\rho B^\sigma+2R^{\rho}{}_{(\mu}B_{\nu)}B_\rho
+\frac{1}{2} g_{\mu\nu}\nabla_\rho\nabla_\sigma(B^\rho B^\sigma)
\nonumber\\&\qquad\qquad
+\frac{1}{2}\Box(B_\mu B_\nu)
-\nabla_\rho\nabla_{(\mu}(B_{\nu)}B^\rho)
\Bigg]
-\frac{\kappa}{2}
\left(
2B_{\mu\lambda}B_{\nu}{}^{\lambda}
-\frac{1}{2}g_{\mu\nu}
B_{\rho\sigma}B^{\rho\sigma}
\right)
\nonumber\\
&\qquad
+2\kappa
\left(
\frac{1}{2}g_{\mu\nu}V
-\frac{\partial V}{\partial X}
B_\mu B_\nu
\right)
-\frac{1}{4}
\left(
2F_{\mu\lambda}F_{\nu}{}^{\lambda}
-\frac{1}{2}g_{\mu\nu}
F_{\rho\sigma}F^{\rho\sigma}
\right)
\nonumber\\
&\qquad
+\gamma_1
\Bigg[
B_\rho B^\rho
\left(
2F_{\mu\lambda}F_{\nu}{}^{\lambda}
-\frac{1}{2}g_{\mu\nu}
F_{\rho\sigma}F^{\rho\sigma}
\right)
+B_\mu B_\nu
F_{\rho\sigma}F^{\rho\sigma}
\Bigg]
\nonumber\\
&\qquad
+\gamma_2
\Bigg[
-\frac{1}{2}g_{\mu\nu}
B^\rho B_\lambda
F_{\rho\sigma}
F^{\lambda\sigma}
+2B_\sigma B_{(\mu}
F_{\nu)\rho}
F^{\sigma\rho}
+B^\rho B^\sigma
F_{\mu\rho}
F_{\nu\sigma}
\Bigg]
=0,\\[5pt]
E_B^{\nu}
\equiv &\;
\gamma R^{\mu\nu}B_\mu
+\kappa\nabla_\mu B^{\mu\nu}
-2\kappa
\frac{\partial V}{\partial X}
B^\nu
+\gamma_1
B^\nu
F_{\rho\sigma}F^{\rho\sigma}
+\gamma_2
B^\mu
F_{\mu\lambda}
F^{\nu\lambda}
=0,\\[5pt]
E_A^{\nu}
\equiv &\;
\nabla_\mu
\left[
\left(
1-4\gamma_1B_\lambda B^\lambda
\right)
F^{\mu\nu}
+4\gamma_2
B_\lambda
F^{\lambda[\mu}
B^{\nu]}\right]
=0,
\end{align}
where $G_{\mu\nu}\equiv R_{\mu\nu}-\frac{1}{2}g_{\mu\nu}R$ is the Einstein 
tensor and $X=B_{\mu}B^\mu \pm b^2$. The solutions of these field equations 
yield bumblebee dyonic Reissner-Nordstr\"om (RN)-like BHs carrying both 
electric and magnetic charges. Following Ref.~\cite{our_BH}, we consider the 
corresponding static and spherically symmetric spacetime solution as 
\begin{equation}
ds^2=-f(r)dt^2+\frac{1}{g(r)}dr^2+r^2d\Omega,
\end{equation}
where $d\Omega=d\theta^2+\sin^2\theta{d\phi^2}$ and
\begin{align}
\label{fr}
f(r) & = 1-\frac{m}{r}+\frac{q^2}{2(2+\xi)r^2}+\frac{(1+\xi)p^2}{2(2+3\xi)r^2},
\\[5pt]
\label{gr}
g(r) & =\frac{f(r)}{1+\xi}.
\end{align}
Here, the parameter $m$ is related to mass, $q$ denotes electric charge, and 
$p$ denotes magnetic charge of the BHs. The Lorentz-violating parameter is 
defined as $\xi=b^2\gamma$ and the coupling constants $\gamma_1$ and 
$\gamma_2$ can be defined as
\begin{equation}
\gamma_1 = \frac{\gamma}{4(2+3\xi)},
\qquad
\gamma_2 = -\frac{2\gamma(1+\xi)}{(2+\xi)(2+3\xi)}.
\end{equation}
It is to be noted that to recover the Schwarzschild solution in its standard 
form for $\xi=q=p=0$ from the metric functions \eqref{fr} and \eqref{gr}, one
may choose $m = 2M$, where $M$ is the mass of the BHs. 

\begin{figure}[!h]
    \centerline{
    \includegraphics[scale=0.48]{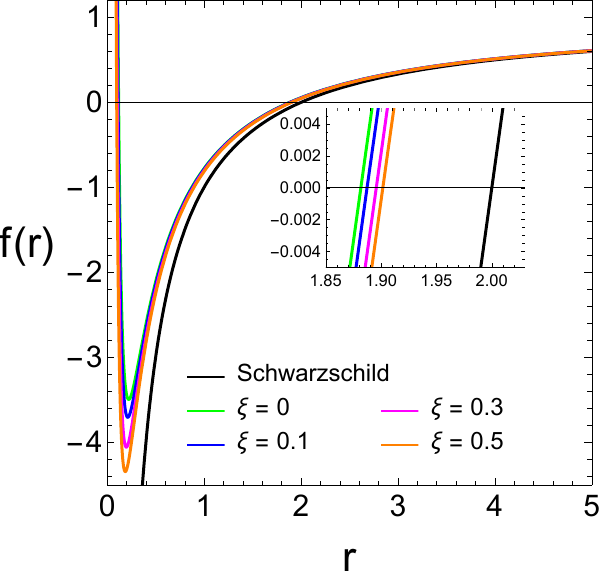}
    \hspace{0.5cm}
    \includegraphics[scale=0.48]{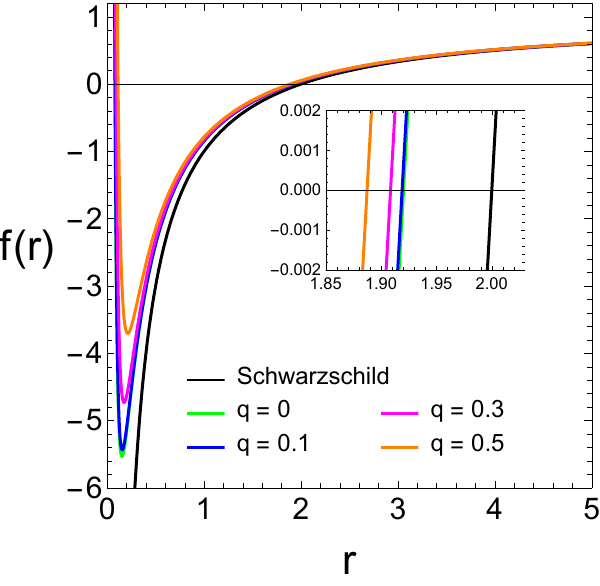}
    \hspace{0.5cm}
    \includegraphics[scale=0.495]{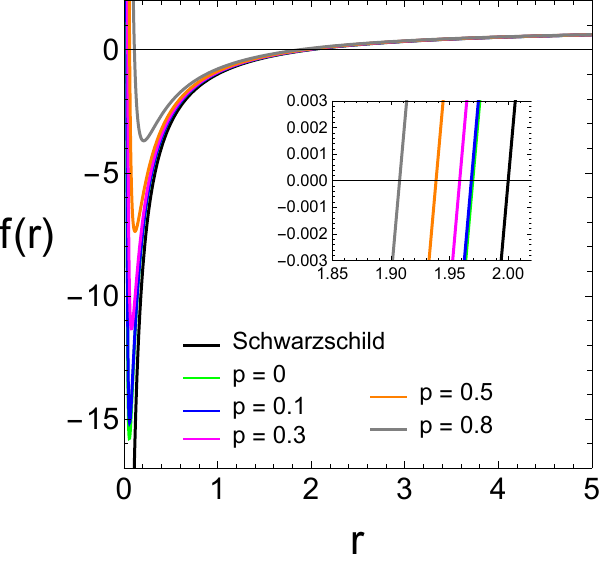}}
    \vspace{0.3cm}
    \centerline{
    \includegraphics[scale=0.48]{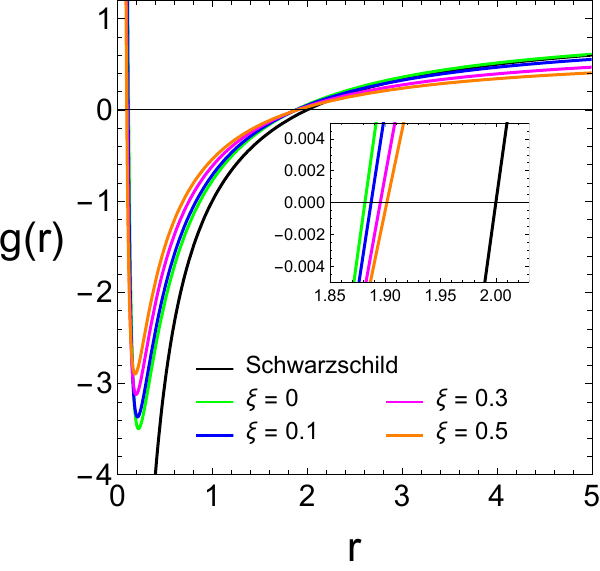}
    \hspace{0.5cm}
    \includegraphics[scale=0.48]{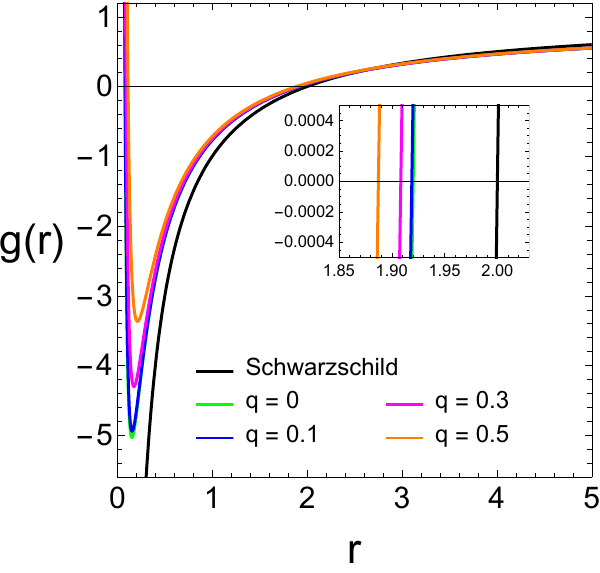}
    \hspace{0.5cm}
    \includegraphics[scale=0.495]{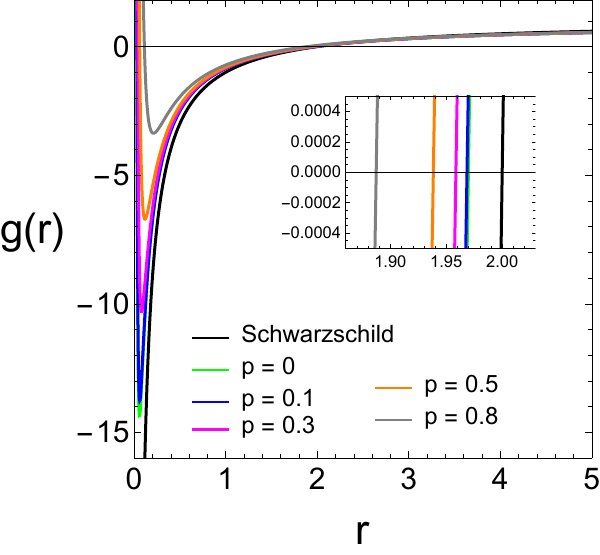}}
    \vspace{-0.1cm}
    \caption{Behaviours of metric functions $f(r)$ and $g(r)$ of the 
considered bumblebee dyonic RN-like BHs with different values of the 
parameters $(\xi, q, p)$. In each plot, the remaining parameters are kept fixed at $\xi=0.1$, $q=0.5$, 
$p=0.8$ as applicable.}
    \label{fig1}
\end{figure}

To understand the geometric properties of the BHs discussed above, we first 
examine the behaviours of the metric functions $f(r)$ and $g(r)$ for 
different values of the parameters $(\xi, p, q)$. Here, we consider 
$m=2$, i.e.~$M=1$ throughout this work unless mentioned otherwise. The zeros 
of $g(r)$, i.e.~roots of $g(r)=0$, correspond to the locations of possible 
horizons of the BHs. In this context, it needs to be mentioned that, 
as is obvious from Eqs.~\eqref{fr} and \eqref{gr}, all zeros of $g(r)$ lead to
zeros of $f(r)$ and vise versa. Thus, all horizons of the BHs are Killing
horizons. In Fig.~\ref{fig1}, the behaviours of metric functions $f(r)$ and 
$g(r)$ as functions of $r$ are shown, which highlight the horizon structure of 
different cases. The upper row shows the behaviour of $f(r)$, and the lower row 
shows that of $g(r)$. The left panels depict the behaviour of metric 
functions for different values of the Lorentz violating parameter $\xi$, 
whereas the middle panels and right panels show their behaviours for different 
values of electric charge parameter $q$ and magnetic charge parameter $p$, 
respectively. 
\section{Quasinormal modes of black holes}
\label{sec3}
We study the QNMs obtained for the considered bumblebee 
dyonic RN-like BHs for two types of perturbations, viz., massless scalar 
perturbations and gravitational perturbations, using the 6th-order Pad\'{e} 
averaged WKB approximation method. For QNMs of the BHs obtained from 
perturbations of the test field, i.e., the scalar field, the effect of the 
field on the spacetime is considered to be negligible. The equations governing 
the perturbations are then formulated on the fixed BH background and can be 
reduced to Schr\"odinger-like wave equations. In the case of the scalar field, 
the corresponding perturbations are governed by the Klein-Gordon equation. In 
contrast, gravitational perturbations involve perturbations on the spacetime 
itself. In this case, both the metric and the gravitational field equations are 
perturbed, leading to Schr\"odinger-like wave equation that describes the 
propagation of the gravitational perturbations. The gravitational 
perturbations can be classified into two sectors: axial (odd-parity) 
perturbations and polar (even-parity) perturbations. In this work, we restrict 
our analysis to axial perturbations only and calculate the QNMs associated 
with them.
\subsection{Scalar Perturbations}
We start by considering a massless scalar field $\Phi$ in the vicinity of the 
BH spacetime. In this case, the equation of motion of the field can be 
described by the Klein-Gordon equation in curved spacetime, as given by 
\cite{scalar_perturbation}
\begin{equation}\label{Klein_Gordon}
\Box{\Phi} = \frac{1}{\sqrt{-g}}\partial_{\mu}\left(\sqrt{-g}\,g^{\mu\nu}\partial_{\nu}\Phi\right)=0.
\end{equation}
For the considered spherically symmetric spacetime, the field $\Phi$ can be 
decomposed as
\begin{equation}\label{scalar_filed}
\Phi(t,r,\theta,\phi)=\frac{1}{r}\sum_{l}\psi_{s}(t,r)\,Y_{lm}(\theta,\phi),
\end{equation}
where $Y_{lm}$ is the spherical harmonic and the function $\psi_{s}(t,r)$ is 
the radial time dependent part of the field. Substituting 
Eq.~\eqref{scalar_filed} in Eq.~\eqref{Klein_Gordon}, we can get the 
Schr\"odinger-like wave equation as 
\begin{equation}\label{wave_equation}
\frac{d^{2}\psi_s}{dr_{*}^{2}}+\left(\omega^{2}-V_{s}(r)\right)\psi_s=0,
\end{equation}
where $\omega$ is the frequency of oscillation of the perturbation, and $r_{*}$ 
is the tortoise coordinate, which can be defined as
\begin{equation}
\frac{dr_{*}}{dr}=\frac{1}{\sqrt{f(r)g(r)}},
\end{equation}
and $V_s(r)$ is the effective potential of the field, given by 
\cite{DJ, scalar_perturbation}
\begin{equation}
V_s(r)=f(r)\!\left[\frac{l(l+1)}{r^2}+\frac{\partial f(r)/\partial r}{r(1+\xi)}\right].
\end{equation}
Here, $l$ is the multipole number. For a BH with asymptotically flat 
spacetime, the Eq.~\eqref{wave_equation} should obey the QNM boundary 
conditions that the ripples of perturbations must be purely ingoing near the 
event horizon and purely outgoing at spatial infinity, i.e., 
\begin{equation}
\psi_s(r_*) \sim
\begin{cases}
A\, e^{+i\omega r_*}, & r_* \rightarrow -\infty, \\[4pt]
B\, e^{-i\omega r_*}, & r_* \rightarrow +\infty,
\end{cases}
\end{equation}
where $A$ and $B$ are amplitudes of ingoing and outgoing waves, respectively.
\subsection{Gravitational Perturbations}
As mentioned already, in this present work, we restrict to axial metric 
perturbations only, and do 
not consider the perturbations of the background bumblebee field and the 
electromagnetic field, i.e., we consider $\delta B_\mu=0$ and 
$\delta A_\mu=0$. The effects of the background bumblebee field and the 
electromagnetic field on perturbations are incorporated through the background 
metric and the resulting effective potential. So, considering the axial 
metric perturbations, the perturbed spacetime metric can be expressed as 
\cite{gravitational_perturbation2}
\begin{equation}
ds^2 = -|g_{tt}|\,dt^2 + r^2\sin^2\theta\left(d\phi-a_1dt-a_2dr-a_3d\theta\right)^2 + g_{rr}\,dr^2 + r^2d\theta^2,
\end{equation}
where $a_1$, $a_2$ and $a_3$ are functions of time $t$, radial coordinate $r$ 
and polar angle $\theta$, respectively, and describe the perturbations of the 
background metric. Whereas $g_{tt}$ and $g_{rr}$ are independent of $t$ and 
$\theta$ and represent the unperturbed background metric. Now, following 
the formalism of gravitational perturbations presented in 
Refs.~\cite{gravitational_perturbation1, gravitational_perturbation2}, the 
axial gravitational perturbations can be described by the equation,
\begin{equation}
R_{(a)(b)}=0,
\end{equation}
where \(R_{(a)(b)}\) denotes the perturbed Ricci tensor in the tetrad 
formalism. The $\theta\phi$ and $r\phi$ components of the above perturbed 
field equation give \cite{gravitational_perturbation2}
\begin{align}\label{phi_theta}
\left[r^2\sqrt{|g_{tt}|g_{rr}^{-1}}\left(a_{2,\theta}-a_{3,r}\right)\right]_{,r}& =\;
r^2\sqrt{|g_{tt}|^{-1}g_{rr}}\left(a_{1,\theta}-a_{3,t}\right)_{,t},\\[5pt]
\label{r_phi}
\left[r^2\sqrt{|g_{tt}|g_{rr}^{-1}}\left(a_{3,r}-a_{2,\theta}\right)\sin^3\theta\right]_{,\theta}
& =\;\frac{r^4\sin^3\theta}{\sqrt{|g_{tt}|g_{rr}}}\left(a_{1,r}-a_{2,t}\right)_{,t}.
\end{align}
Here, the radial and angular dependence can be separated using the 
decomposition of the function,
\begin{equation}
\mathcal{F}_g(r,\theta)=\mathcal{F}_g(r)Y(\theta),
\end{equation}
where the angular function $Y(\theta)$ satisfies the equation,
\begin{equation}
\frac{d}{d\theta}\left(\sin^{-3}\theta\frac{dY}{d\theta}\right) = -\left[l(l+1)-2\right]Y\sin^{-3}\theta,
\end{equation}
and $\mathcal{F}_g(r)$ is the radial function. Using this decomposition in 
Eqs.~\eqref{phi_theta} and \eqref{r_phi}, and simplifying, we can get the 
Schr\"odinger-like equation, as given by 
\begin{equation}
\frac{d^{2}\psi_g}{dr_{*}^{2}}+\left(\omega^{2}-V_{g}(r)\right)\psi_g=0,
\end{equation}
where $\psi_g =\mathcal{F}_g/r$ and $V_g$ is the effective potential 
associated with gravitational field perturbations and is given by 
\cite{grav_eff_potential},
\begin{equation}
V_g(r) = f(r)\!\left[
\frac{(l-1)(l+2)}{r^{2}} +\frac{2f(r)}{r^{2}(1+\xi)}-\frac{\partial f(r)/\partial r}{r(1+\xi)}\right].
\end{equation}

\begin{figure}[!h]
    \centerline{
    \includegraphics[scale=0.62]{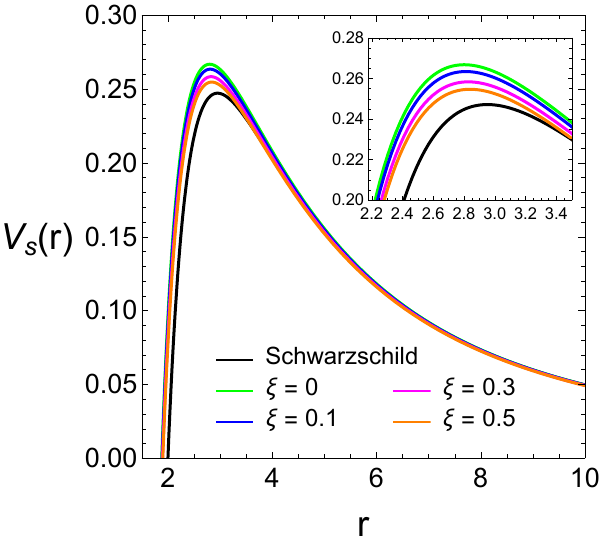}
    \hspace{0.5cm}
    \includegraphics[scale=0.62]{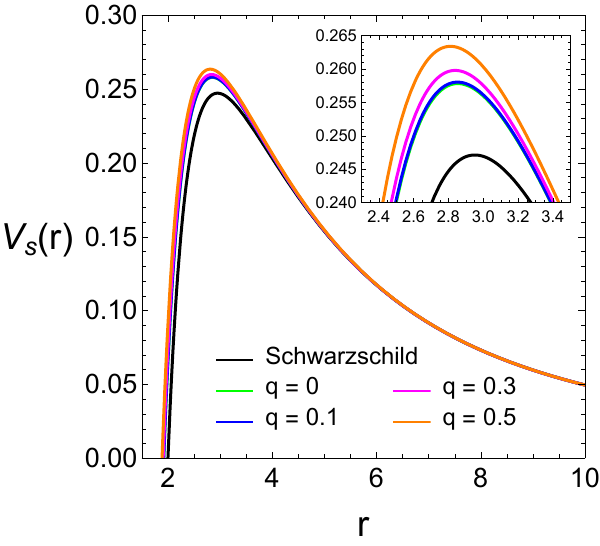}}
    \vspace{0.3cm}
    \centerline{
    \includegraphics[scale=0.62]{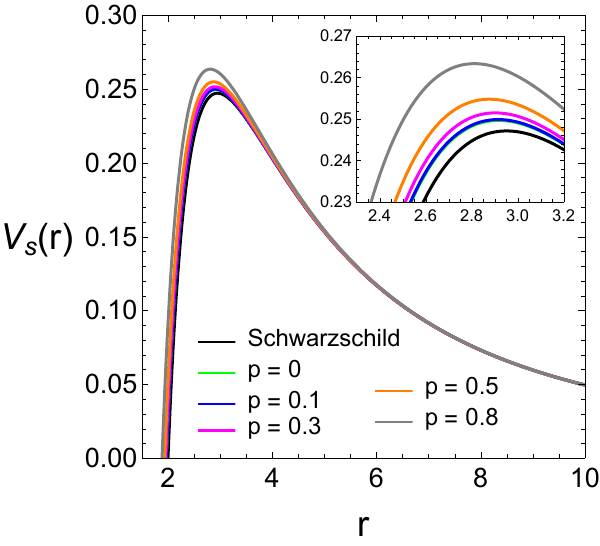}
    \hspace{0.5cm}
    \includegraphics[scale=0.6]{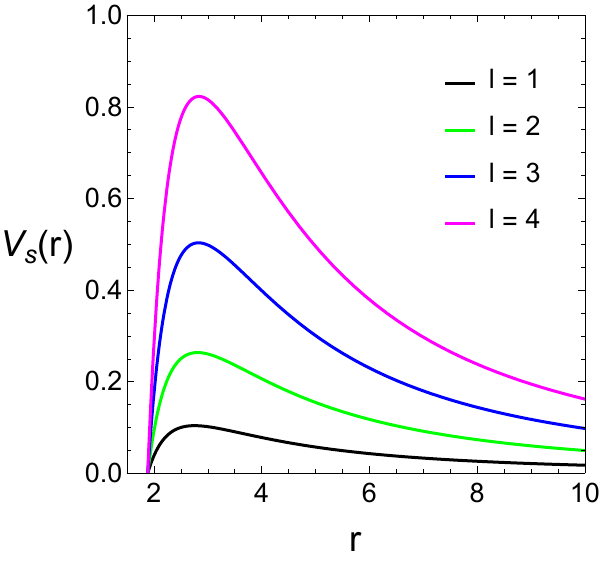}}
    \vspace{-0.2cm}
    \caption{Variations of scalar effective potential $V_s(r)$ as a function 
of radial distance $r$ for different values of $\xi$, $p$, $q$, and $l$. In
each plot, the remaining parameters are kept fixed at $\xi=0.1$, $q=0.5$, 
$p=0.8$ and $l=2$ as applicable.}
    \label{fig2}
\end{figure}
\begin{figure}[!h]
    \centerline{
    \includegraphics[scale=0.625]{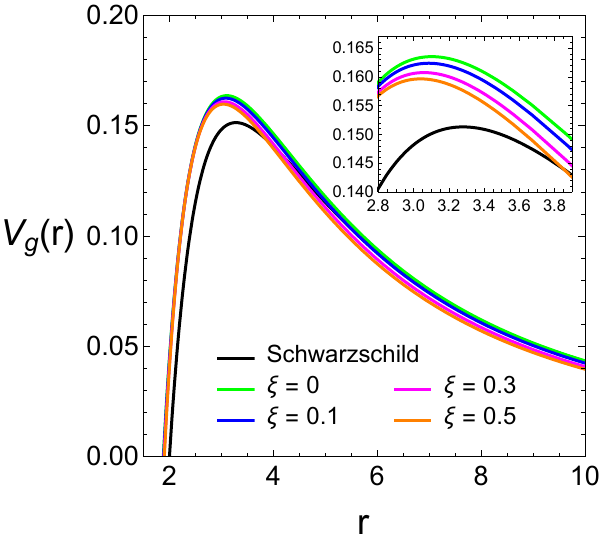}
    \hspace{1cm}
    \includegraphics[scale=0.625]{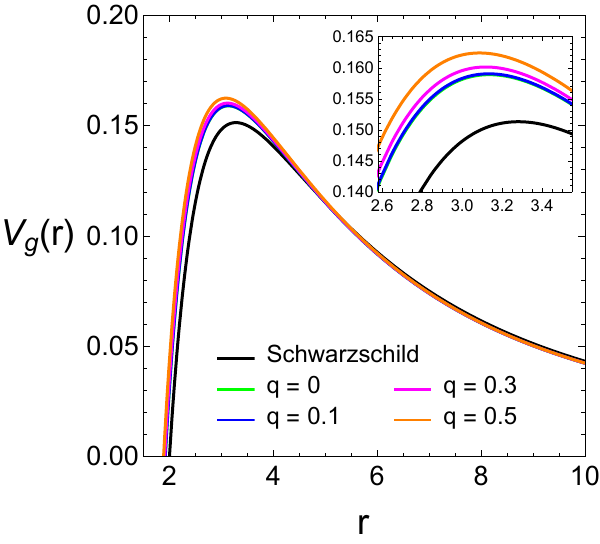}}
    \vspace{0.3cm}
    \centerline{
    \includegraphics[scale=0.625]{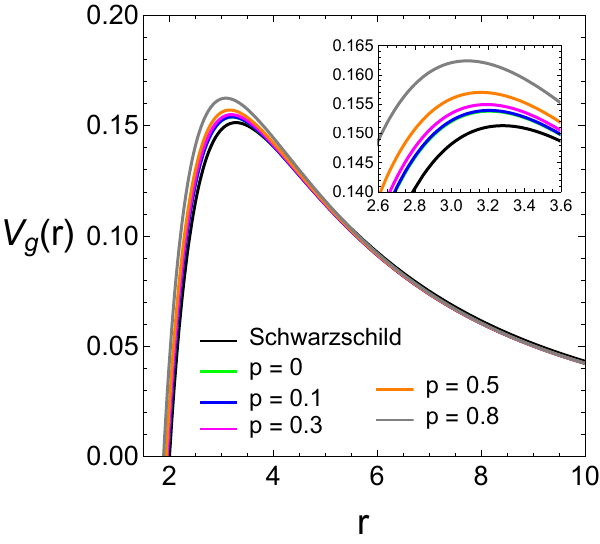}
    \hspace{1cm}
    \includegraphics[scale=0.6]{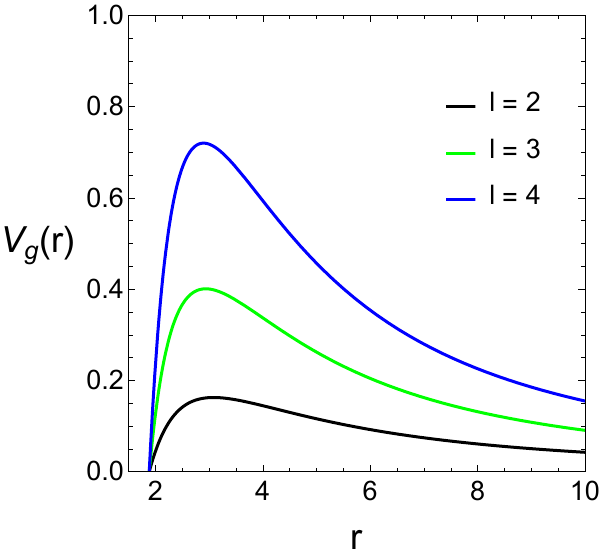}}
    \vspace{-0.2cm}
    \caption{Variations of effective potential $V_g(r)$ of gravitational 
perturbations as a function of radial distance $r$ for different values of 
$\xi$, $p$, $q$, $l$. In each plot, the remaining parameters are kept fixed at 
$\xi=0.1$, $q=0.5$, $p=0.8$ and $l=2$, as applicable.}
    \label{fig3}
\end{figure}
We study the behaviours of the effective potentials $V_s(r)$ and 
$V_g(r)$ as functions of radial distance $r$ for different values of the 
model parameters as well as the multipole number. Fig.~\ref{fig2} shows the 
effective potential $V_s(r)$ of the scalar field plotted for various values of 
the said parameters. Similarly, in Fig.~\ref{fig3}, the effective potential 
associated with gravitational perturbations, i.e., $V_g(r)$ is plotted for 
various values of those parameters. In each plot, one parameter is varied, and 
the other parameters are fixed at $\xi=0.1$, $q=0.5$, $p=0.8$  and $l=2$, as 
applicable. From the figures it is seen that each potential attains a single 
positive peak outside the event horizon and approaches zero as  
$r\rightarrow \infty$. The Lorentz violating parameter $\xi$ has a significant 
effect on the height of the potential barrier. For $\xi=0$, the peak of the 
potential attains the maximum value and as $\xi$ increases the height of the 
peak decreases. On the other hand, with an increase in electric charge $q$ and 
magnetic charge $p$, the maximum value of the effective potential increases. 
Again, the effect of the multipole number on the effective potential is the 
strongest. As $l$ increases, the height of the potential barrier also 
increases significantly. However, the gravitational perturbation 
potential is not well defined for the $l=1$ value, and hence this value of $l$ 
is not considered in the study of the gravitational perturbations.
\subsection{Calculation of Quasinormal Modes using WKB Approximation Method 
with Pad\'{e} average}
\begin{figure}[!h]
    \centerline{
    \includegraphics[scale=0.5]{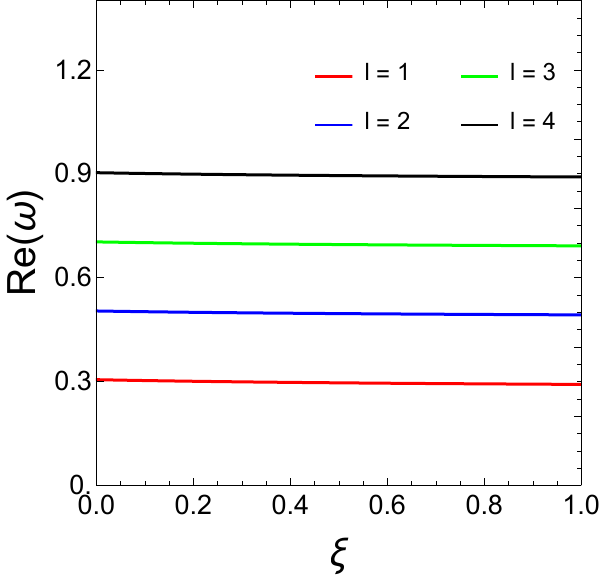}
    \hspace{0.5cm}
    \includegraphics[scale=0.5]{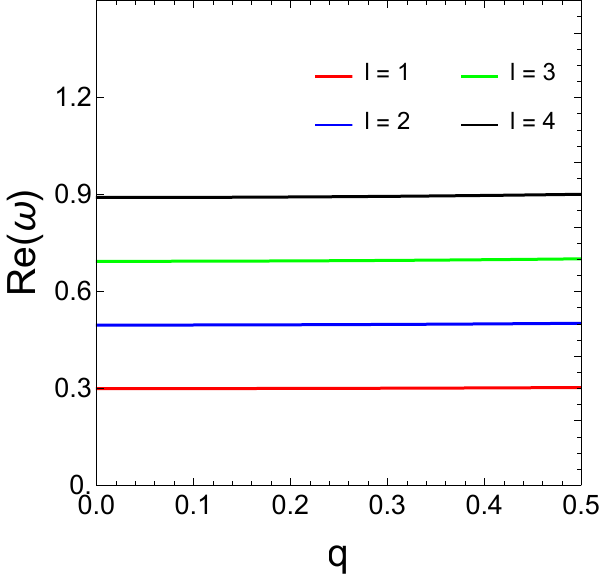}
    \hspace{0.5cm}
    \includegraphics[scale=0.5]{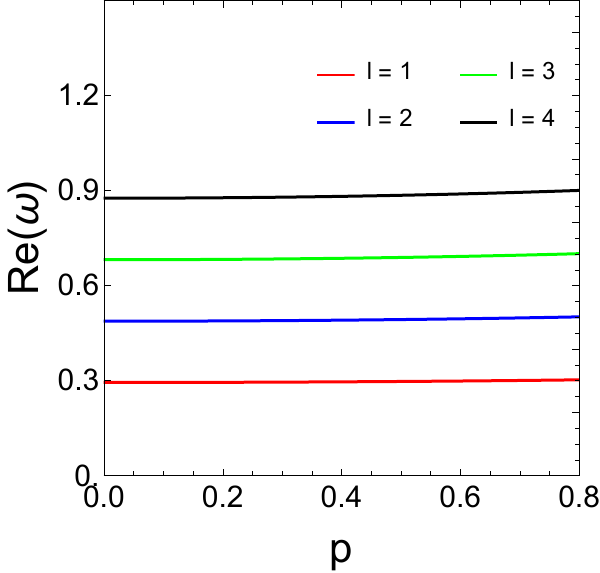}}
    \vspace{-0.2cm}
    \caption{Variation of the real part of QNMs for scalar perturbations with
respect to different values of the parameters $\xi$, $q$ and $p$. In each plot the remaining parameters are kept fixed at $\xi=0.1$, $q=0.5$ and $p=0.8$ as applicable.}
    \label{fig4}
\end{figure}

\begin{figure}[!h]
    \centerline{
    \includegraphics[scale=0.55]{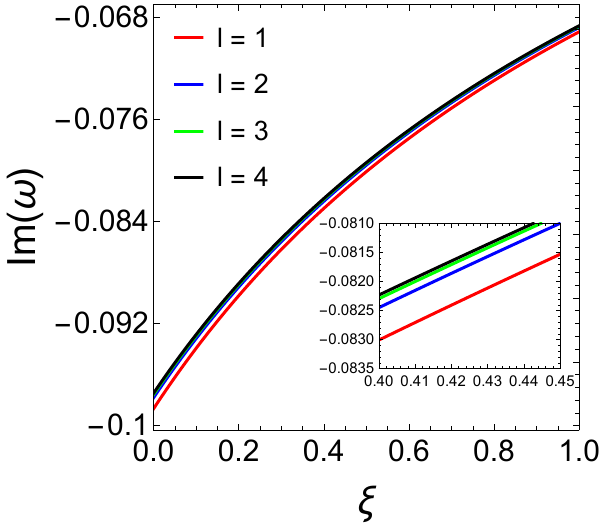}
    \hspace{0.2cm}
    \includegraphics[scale=0.55]{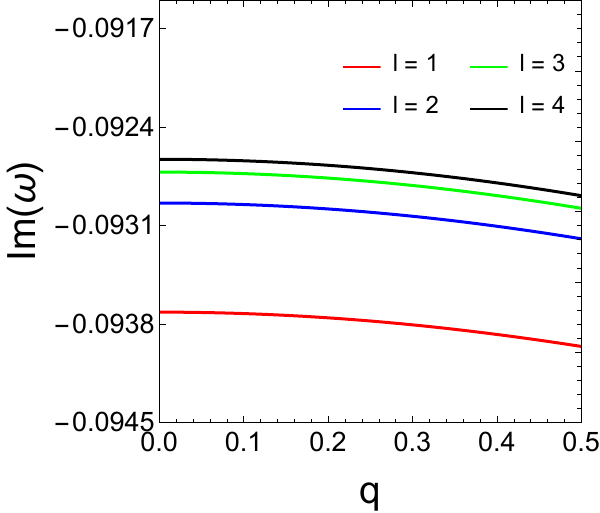}
    \hspace{0.2cm}
    \includegraphics[scale=0.55]{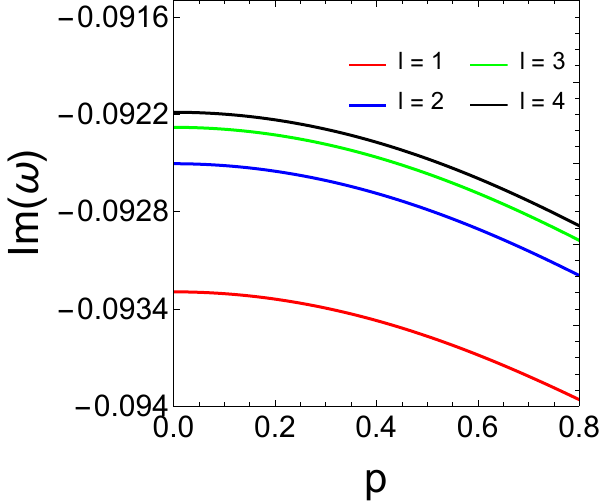}}
    \vspace{-0.2cm}
    \caption{Variation of imaginary part of QNMs for scalar perturbations with
respect to different values of the parameters $\xi$, $q$ and $p$. In each plot the remaining parameters are kept fixed at $\xi=0.1$, $q=0.5$ and $p=0.8$ as applicable.}
    \label{fig5}
\end{figure}

\begin{figure}[!h]
    \centerline{
    \includegraphics[scale=0.5]{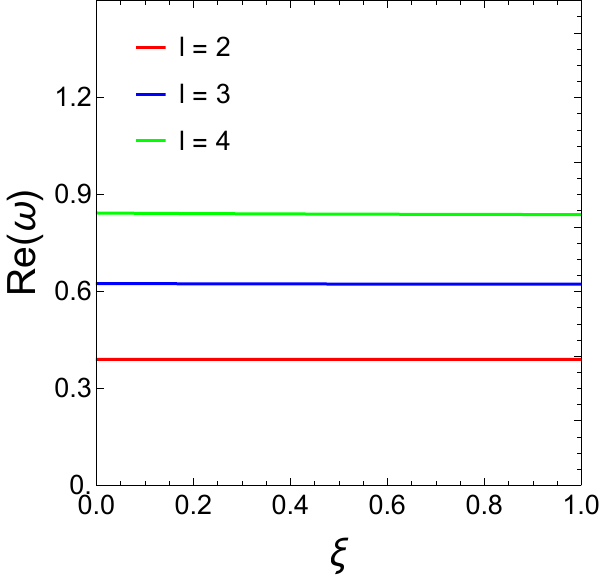}
    \hspace{0.5cm}
    \includegraphics[scale=0.5]{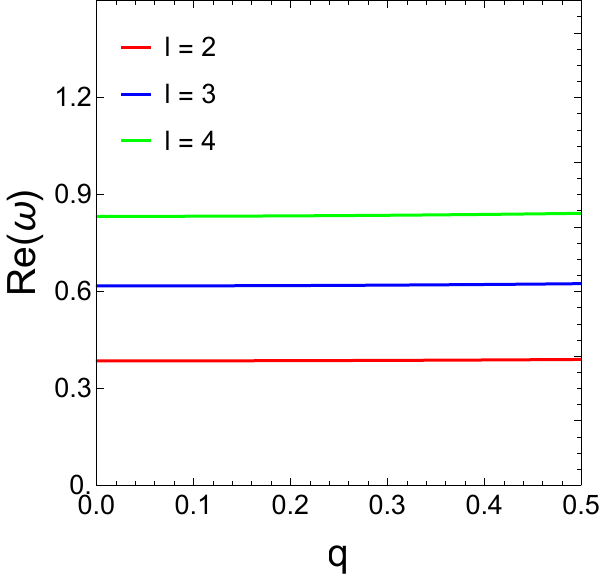}
    \hspace{0.5cm}
    \includegraphics[scale=0.5]{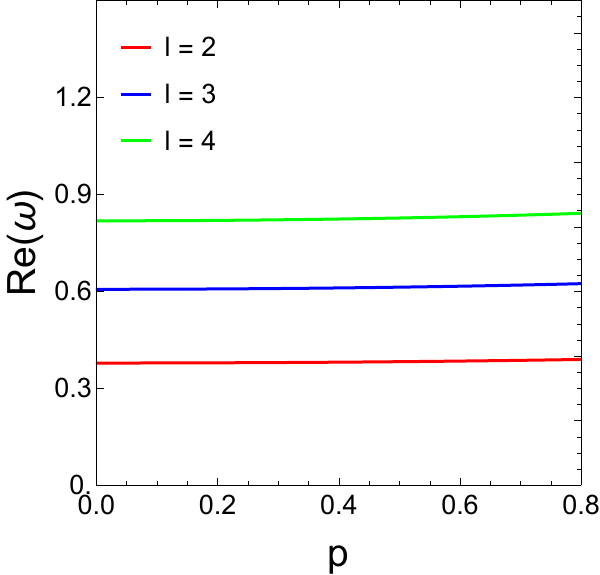}}
    \vspace{-0.2cm}
    \caption{Variation of real part of QNMs for gravitational perturbations 
with respect to different values of the parameters $\xi$, $q$ and $p$. In each plot the remaining parameters are kept fixed at $\xi=0.1$, $q=0.5$ and $p=0.8$ as applicable.}
    \label{fig6}
\end{figure}

\begin{figure}[!h]
    \centerline{
    \includegraphics[scale=0.55]{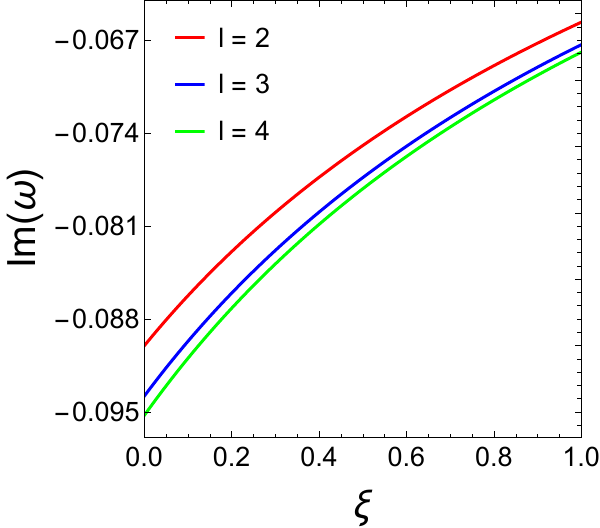}
    \hspace{0.2cm}
    \includegraphics[scale=0.55]{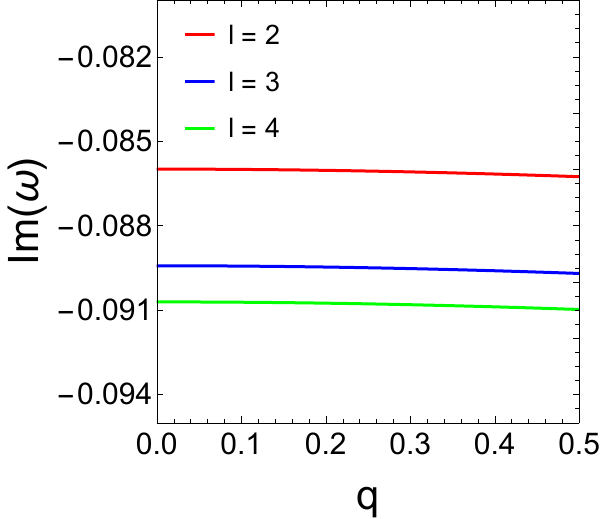}
    \hspace{0.2cm}
    \includegraphics[scale=0.55]{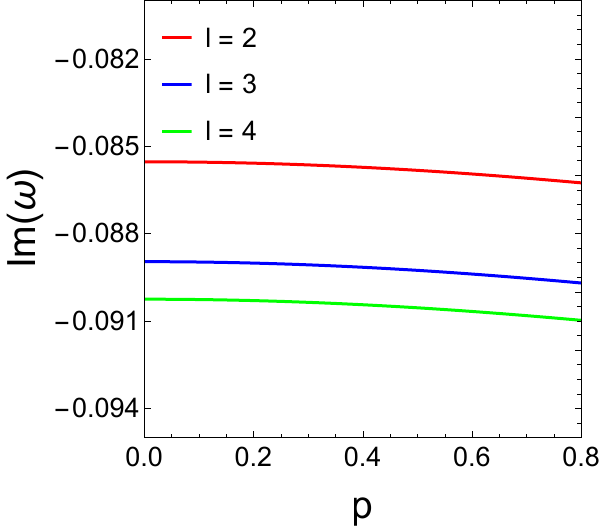}}
    \vspace{-0.2cm}
    \caption{Variation of imaginary part of QNMs of gravitational 
perturbations with respect to different values of the parameters $\xi$, $q$ 
and $p$. In each plot the remaining parameters are kept fixed at $\xi=0.1$, $q=0.5$ and $p=0.8$ as applicable.}
    \label{fig7}
\end{figure}

We calculate the QNMs of the BHs defined in the preceding sections using 
higher order WKB approximation method. The WKB method to calculate the QNMs 
was first suggested by Schutz and Will in Ref.~\cite{WKB2}. Subsequently, the 
method was extended to higher orders \cite{WKB3, WKB4, WKB5}. Later, in 
Ref.~\cite{WKB5}, the implementation of Pad\'e approximation to the WKB method 
was proposed, and from subsequent studies it was seen that the method improves 
the results of QNMs with a higher accuracy \cite{WKB4}. In this work, we use 
the Pad\'e averaged 6th-order WKB approximation method to calculate the QNMs 
associated with the BHs under consideration.\\
\indent The numerical values of the QNMs for scalar and gravitational 
perturbations for different values of $l$ and $\xi$ with fixed values of 
$q=0.5$ and $p=0.8$, obtained from Pad\'{e} averaged 6th-order WKB 
approximation method, are shown in Tables \ref{table1} and \ref{table2}, 
respectively. The error associated with the QNM calculations using the WKB 
approximation has been estimated using the formula \cite{WKB4},
\begin{equation}
\Delta_6 = \frac{|\omega_7 - \omega_5|}{2},
\end{equation}
where $\omega_7$ and $\omega_5$ are the QNMs obtained from 7th and 5th order 
Pad\'{e} averaged WKB method. From the tables, it is seen that for both
scalar and gravitational perturbations, the oscillation frequency increases with
increasing multipole number $l$, whereas the damping decreases with increasing
$l$.  
\begin{table}[ht]
\centering
\caption{Pad\'e averaged 6th-order WKB quasinormal frequencies of the BHs for 
different values of $\xi$ and multipole number $l$, with $n=0, p=0.8$ and 
$q=0.5$ for scalar perturbations.}
\vspace{8pt}
\label{table1}
\begin{tabular}{cccc}
\toprule
Multipole & $\xi$ & Pad\'e averaged 6th-order WKB &
$\Delta_6$ \\
\midrule
Schwarzschild ($l=1$)
& 0 & $0.292910 - 0.0977616i$ & $9.83688\times10^{-5}$ \\
\multirow{3}{*}{$l=1$}
& 0.1 & $0.302411 - 0.0940242i$ & $7.77801\times10^{-5}$ \\
& 0.3 & $0.298600 - 0.0862709i$ & $4.84992\times10^{-5}$ \\
& 0.5 & $0.295819 - 0.0801591i$ & $3.46917\times10^{-5}$ \\

\addlinespace
Schwarzschild ($l=2$)
& 0 & $0.483642 - 0.0967661i$ & $8.51824\times10^{-6}$ \\
\multirow{3}{*}{$l=2$}
& 0.1 & $0.501177 - 0.0931996i$ & $6.69572\times10^{-6}$ \\
& 0.3 & $0.498007 - 0.0856127i$ & $4.05062\times10^{-6}$ \\
& 0.5 & $0.495713 - 0.0796196i$ & $2.56744\times10^{-6}$ \\

\addlinespace
Schwarzschild ($l=3$)
& 0 & $0.675366 - 0.0965006i$ & $1.33295\times10^{-6}$ \\
\multirow{3}{*}{$l=3$}
& 0.1 & $0.700578 - 0.0929785i$ & $1.15201\times10^{-6}$ \\
& 0.3 & $0.697362 - 0.0854389i$ & $6.88674\times10^{-7}$ \\
& 0.5 & $0.695045 - 0.0794783i$ & $4.38446\times10^{-7}$ \\

\addlinespace
Schwarzschild ($l=4$)
& 0 & $0.867416 - 0.0963919i$ & $3.42258\times10^{-7}$ \\
\multirow{3}{*}{$l=4$}
& 0.1 & $0.900182 - 0.0928882i$ & $3.14303\times10^{-7}$ \\
& 0.3 & $0.896692 - 0.0853680i$ & $1.86702\times10^{-7}$ \\
& 0.5 & $0.894186 - 0.0794207i$ & $1.18567\times10^{-7}$ \\
\bottomrule
\end{tabular}
\end{table}
\begin{table}[ht]
\centering
\caption{Pad\'e averaged 6th-order WKB quasinormal frequencies of the BHs for 
different values of $\xi$ and multipole number $l$, with $n=0, p=0.8$ and 
$q=0.5$ for gravitational perturbations.}
\vspace{8pt}
\label{table2}
\begin{tabular}{cccc}
\toprule
Multipole & $\xi$ & Pad\'e averaged 6th-order WKB & $\Delta_6$ \\
\midrule
Schwarzschild ($l=2$)
& 0 & $0.373619 - 0.0888910i$ & $7.35373\times10^{-5}$ \\
\multirow{3}{*}{$l=2$}
& 0.1 & $0.389264 - 0.0862333i$ & $3.37801\times10^{-5}$ \\
& 0.3 & $0.389204 - 0.0799750i$ & $1.56118\times10^{-5}$ \\
& 0.5 & $0.389294 - 0.0749379i$ & $7.15138\times10^{-6}$ \\

\addlinespace
Schwarzschild ($l=3$)
& 0 & $0.599443 - 0.0927025i$ & $1.07703\times10^{-6}$ \\
\multirow{3}{*}{$l=3$}
& 0.1 & $0.623696 - 0.0897004i$ & $4.82898\times10^{-7}$ \\
& 0.3 & $0.623043 - 0.0828466i$ & $3.66044\times10^{-7}$ \\
& 0.5 & $0.622637 - 0.0773626i$ & $2.89319\times10^{-7}$ \\

\addlinespace
Schwarzschild ($l=4$)
& 0 & $0.809178 - 0.0941641i$ & $9.22585\times10^{-8}$ \\
\multirow{3}{*}{$l=4$}
& 0.1 & $0.841309 - 0.0909804i$ & $2.46898\times10^{-7}$ \\
& 0.3 & $0.839887 - 0.0838695i$ & $1.54724\times10^{-7}$ \\
& 0.5 & $0.838908 - 0.0782037i$ & $1.02556\times10^{-7}$ \\

\bottomrule
\end{tabular}
\end{table}
In Fig.~\ref{fig4} and \ref{fig5}, we plot the variations of real part 
and imaginary part of QNMs for scalar perturbations with respect to different 
values of the parameters $(\xi, q, p)$, respectively. Similarly, in 
Fig.~\ref{fig6} and \ref{fig7}, we show how the real part and imaginary part 
of the QNMs associated with gravitational perturbations vary with respect
to the parameters $(\xi, q, p)$, respectively. From the figures, it is evident 
that with increasing Lorentz violating parameter $\xi$, the oscillation 
frequency decreases slightly for all values of $l$ in the case of both scalar 
and gravitational perturbations (same is also evident from Tables \ref{table1} 
and \ref{table2}). Whereas, with increasing electric charge $q$ 
and magnetic charge $p$, the oscillation frequency slightly increases in the 
case of both scalar and gravitational perturbations. In case of the damping 
rate, with increasing $\xi$, the $Im(\omega)$ becomes less negative for both 
scalar and gravitational perturbations, indicating that the damping rate 
decreases and the perturbations become relatively long lived. However, with 
increasing $q$ and $p$, the damping rate increases for both perturbations. 
Furthermore, from the figures, we can see that the damping rate of the scalar 
perturbation is higher for $l=1$ and gradually decreases with increasing $l$, 
whereas in the case of gravitational perturbations, it increases with 
increasing $l$.
\section{Evolution of Scalar and Gravitational perturbations}\label{sec4}
In this section, we investigate the time-domain evolution of scalar and 
gravitational perturbations around the BH spacetime using the time-domain 
integration method described in Ref.~\cite{time_profile1}. By discretizing 
the tortoise coordinate and time as $r_*=i\Delta r_*$ and $t=j\Delta t$, we 
can write $\psi(r_*,t)=\psi(i\Delta r_*,j\Delta t) \equiv \psi_{i,j}$ and 
express the effective potential as $V(r(r_*))=V(r_*,t) \equiv V_{i,j}$. With 
these definitions, the Eq.~\eqref{Klein_Gordon} can be expressed as
\begin{equation}
\frac{\psi_{i+1,j}-2\psi_{i,j}+\psi_{i-1,j}}{\Delta r_*^2}-\frac{\psi_{i,j+1}-2\psi_{i,j}+\psi_{i,j-1}}{\Delta t^2}-V_i\psi_{i,j}=0.
\end{equation}
We impose a Gaussian wave packet as the initial perturbation of the form 
$\psi(r_*,t)=\exp\left[-\frac{(r_*-\mu)^2}{2\sigma^2}\right]$ together with 
the condition $\left.\psi(r_*,t)\right|_{t<0}=0$, where $\mu$ specifies the 
median and $\sigma$ determines the width of the initial wave packet. Thus, 
using these conditions, the time evolution of scalar field perturbations 
can be expressed as 
\begin{equation}
\psi_{i,j+1}=-\psi_{i,j-1}+\left(\frac{\Delta t}{\Delta r_*}\right)^2
\left(\psi_{i+1,j}+\psi_{i-1,j}\right)+\left[2-2\left(\frac{\Delta t}{\Delta r_*}\right)^2-V_i\Delta t^2
\right]\psi_{i,j}.
\end{equation}
In this numerical procedure, the time and spatial steps are chosen such that 
$\frac{\Delta t}{\Delta r_*}<1$ to satisfy the Von Neumann stability 
condition. The same finite difference procedure is applied to gravitational 
perturbations also to obtain their corresponding time profile.\\
\begin{figure}[!h]
    \centerline{
    \includegraphics[scale=0.43]{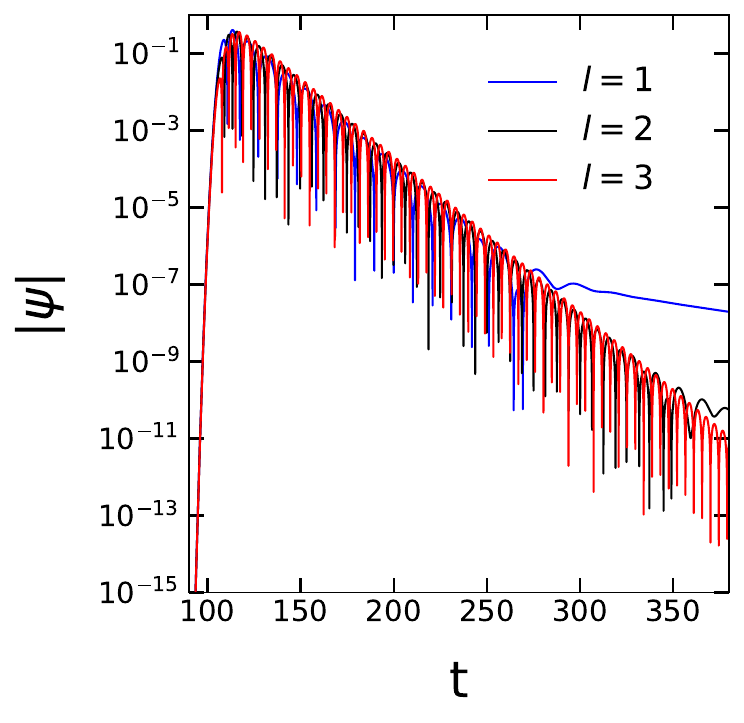}
    \hspace{0.3cm}
    \includegraphics[scale=0.42]{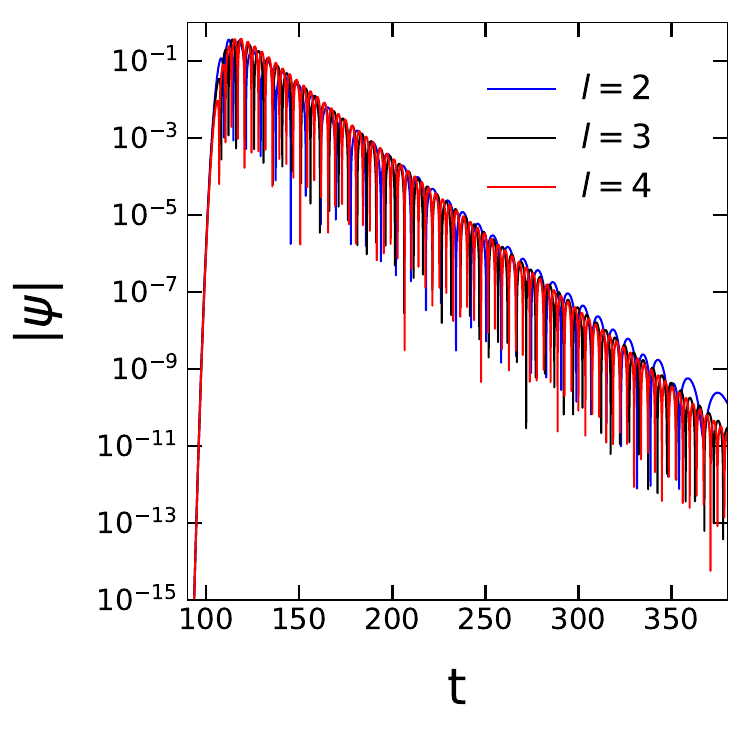}
    \hspace{0.3cm}
    \includegraphics[scale=0.42]{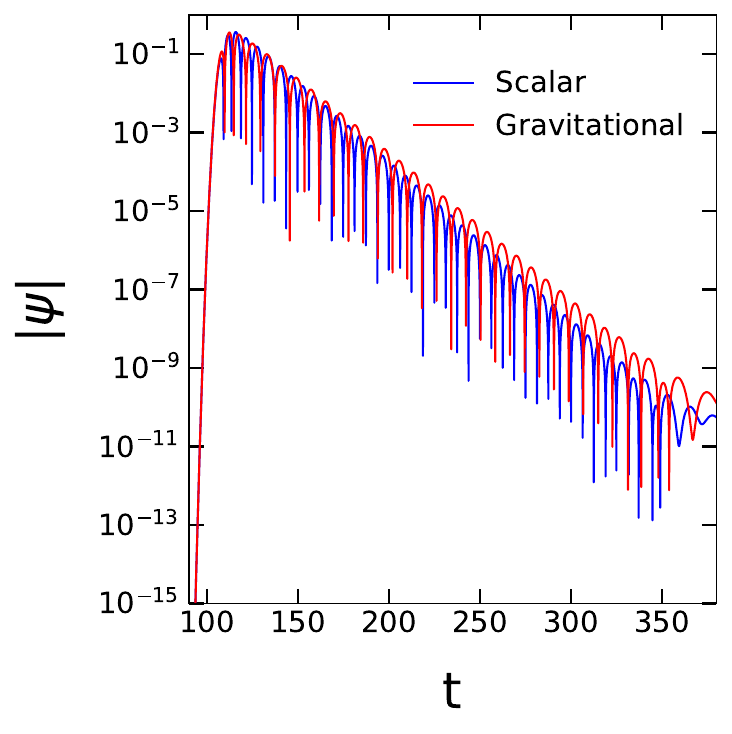}}
    \vspace{-0.2cm}
    \caption{Time-domain profile with $n=0$, $\xi=0.1$, $q=0.5$ and $p=0.8$ 
for different values of $l$. The left panel shows the time profile of scalar 
perturbations, the middle panel shows the time profile of gravitational 
perturbations, and the right panel shows the comparison of the time profile 
of scalar and gravitational perturbations for $l=2$.}
    \label{fig8}
\end{figure}
\indent In Fig.~\ref{fig8}, we plot the time-domain profile of scalar and 
gravitational perturbations considering the values of the parameters as 
$\xi=0.1$, $q=0.5$ and $p=0.8$. The left panel shows the time-domain evolution 
of scalar perturbations for $l=1$, $2$, $3$, the middle panel shows the time 
evolution of gravitational perturbations for $l=2$, $3$, $4$, and the right 
panel shows the comparison between time profiles of the scalar and 
gravitational perturbations for $l=2$. From the figure, it is seen that the 
time-domain profile varies with multipole number in the case of both scalar and 
gravitational perturbations. Also, the oscillation frequency of gravitational 
perturbations is slightly less than that of scalar perturbations, but the 
scalar perturbations decay faster than those of gravitational perturbations. 
These features are in good agreement with the results obtained using 6th-order 
Pad\'e averaged WKB approximation.\\
\indent Furthermore, we employ the Matrix Pencil (MP) method described in 
Refs.~\cite{MP1, MP2, MP3} to extract the QNM frequencies from the time 
profiles. Starting from the time domain waveform, two time-shifted Hankel 
matrices $Y_0$ and $Y_1$ are constructed. We then perform a singular value 
decomposition on $Y_1$ as $Y_1=USV^\dagger$, where $U$ and $V$ are unitary 
matrices of left and right singular vectors, respectively, and $S$ is the 
diagonal matrix containing the singular values. Then the first dominant 
singular values, say $x$ are kept, while the remaining components, assumed to 
be dominated by noise, are discarded. Then the corresponding reduced matrices 
$U_x$, $V_x$ and $S_x$ are used to construct the reduced matrix as 
$Z_x=S^{-1}_xU^{\dagger}_xY_0V_x$. The eigenvalues of $Z_x$ provide estimates 
of the inverse poles associated with the damped oscillatory components of the 
waveform. From these eigenvalues, the complex QNM frequencies were obtained 
following the prescription given in Ref.~\cite{MP3}, with the real and 
imaginary parts describing the oscillation frequency and damping rate, 
respectively. The extracted QNMs were subsequently used to reconstruct the 
time-domain waveform, which was compared with the original numerical time-domain
 profile. The agreement between the original and reconstructed waveforms 
was quantified using the coefficient of determination $R^2$, as given by
\begin{equation}
R^2 =1 -\frac{\displaystyle \sum_i\left(\psi_{\mathrm{Time-Profile}}
-\psi_{\mathrm{MP}}\right)^2}{\displaystyle \sum_i\left(
\psi_{\mathrm{Time-Profile}}-\overline{\psi}_{\mathrm{Time-Profile}}
\right)^2},
\end{equation}
where $\psi_{\mathrm{Time-Profile}}, \psi_{\mathrm{MP}}$ and 
$\overline{\psi}_{\mathrm{Time-Profile}}$ are the amplitudes of the original 
time domain waveform, the waveform reconstructed using the MP method and the
mean value of the original time-domain waveform, respectively.\\
\indent Fig.~\ref{fig9} shows the comparison between the waveforms constructed 
using the MP method and the original time-domain profiles for different values 
of $l$ with $\xi=0.1, q=0.5, p=0.8$ for scalar perturbations. Similarly, 
Fig.~\ref{fig10} illustrates the comparison between the waveforms of 
the time-domain profile and those obtained from the MP method for gravitational 
perturbations. From the figures, it is seen that the waveforms reconstructed 
from QNMs extracted using the MP method are in good agreement with the original 
time-domain waveforms in the case of both scalar and gravitational 
perturbations. 
\begin{figure}[!h]
    \centerline{
    \includegraphics[scale=0.38]{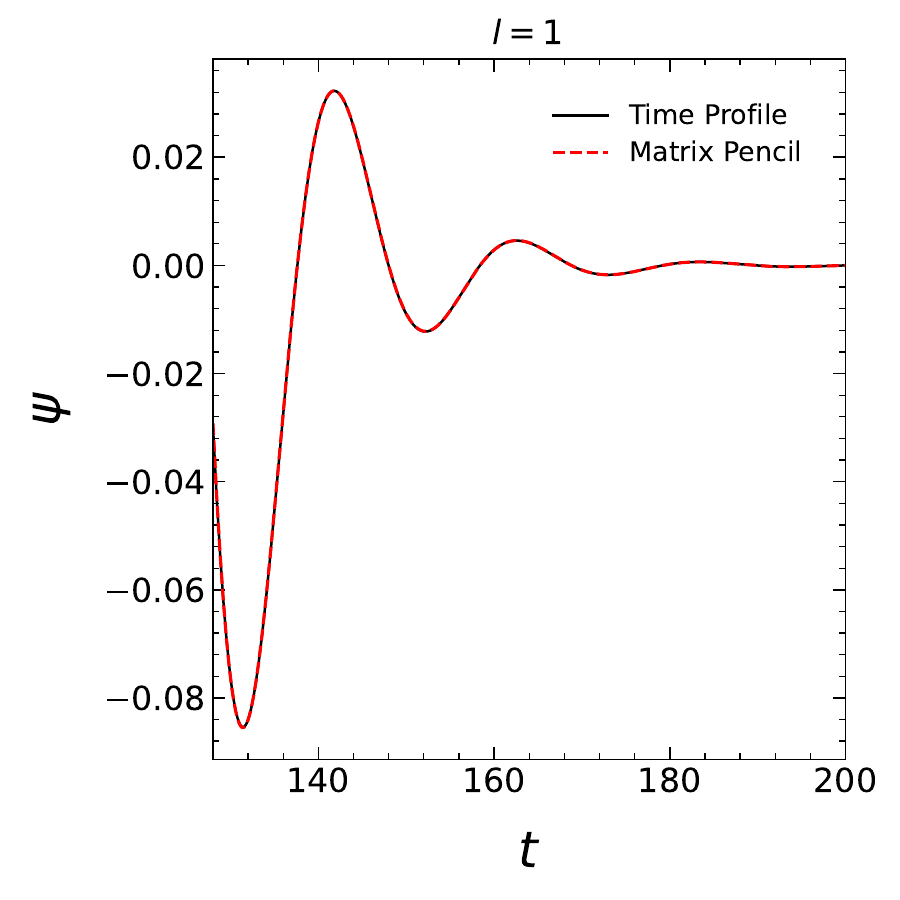}
    \hspace{0.0cm}
    \includegraphics[scale=0.38]{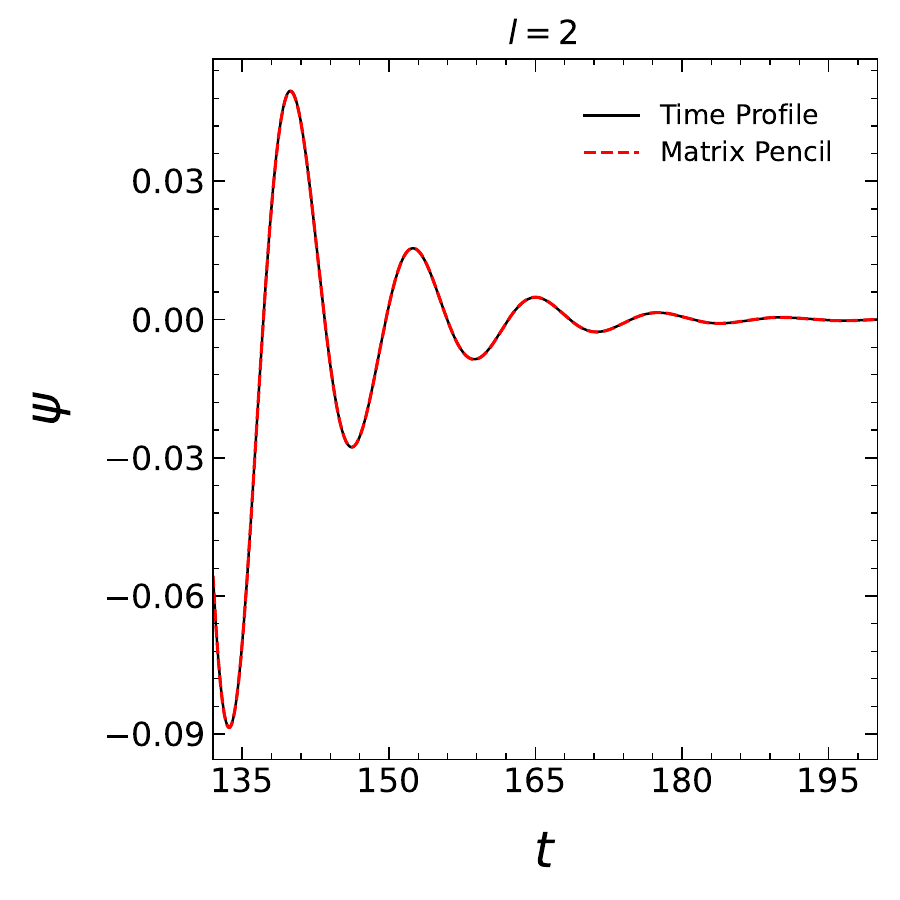}
    \hspace{0.0cm}
    \includegraphics[scale=0.38]{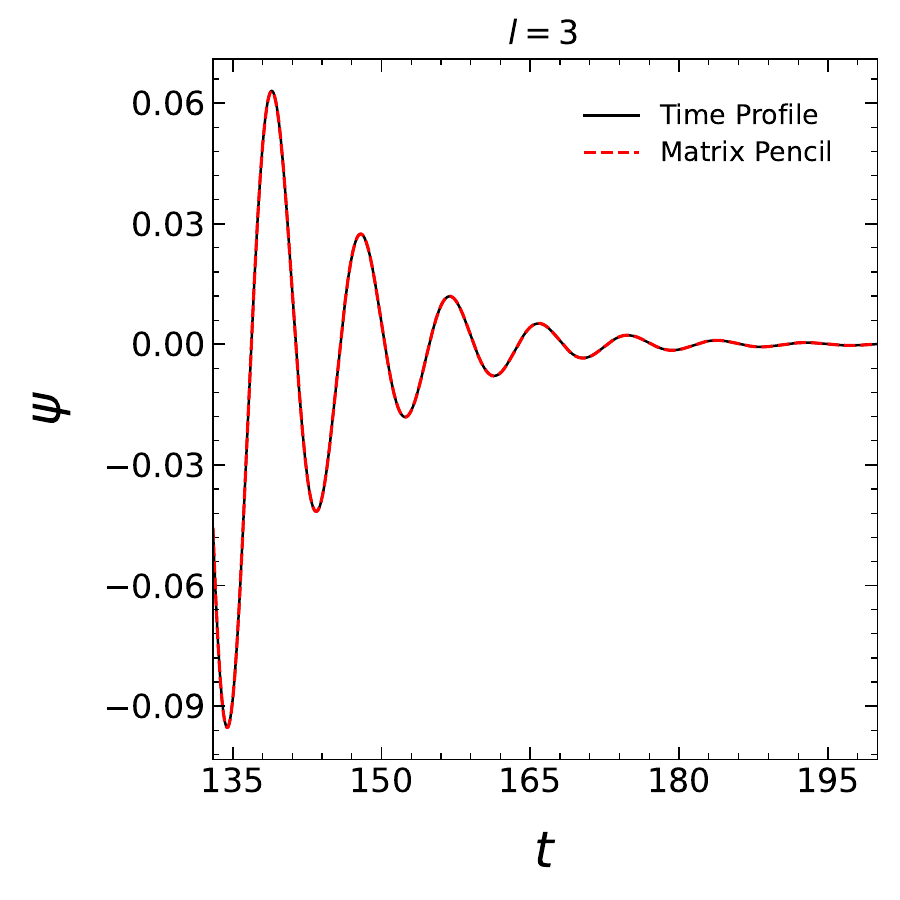}}
    \vspace{-0.2cm}
    \caption{Comparison between the waveforms of the time-domain profile and 
that reconstructed using the Matrix Pencil method with $n=0$, $\xi=0.1$, 
$q=0.5$, $p=0.8$ for different values of $l$ for scalar perturbations.}
    \label{fig9}
\end{figure}
\begin{figure}[!h]
    \centerline{
    \includegraphics[scale=0.38]{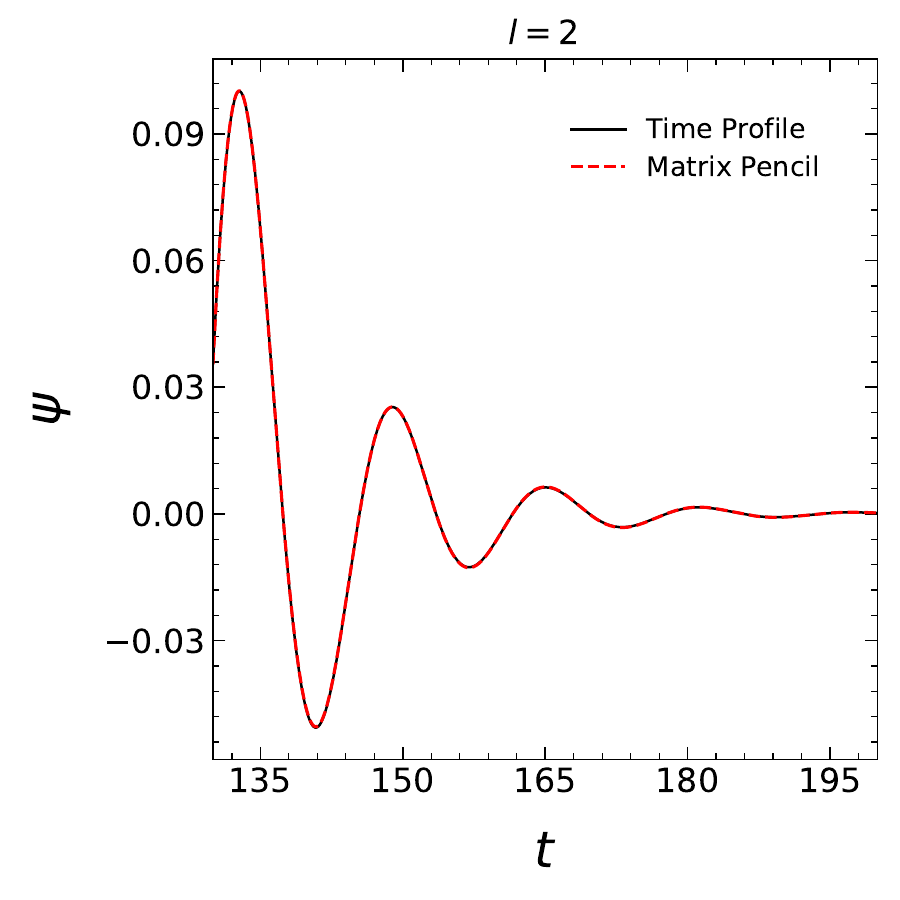}
    \hspace{0.0cm}
    \includegraphics[scale=0.38]{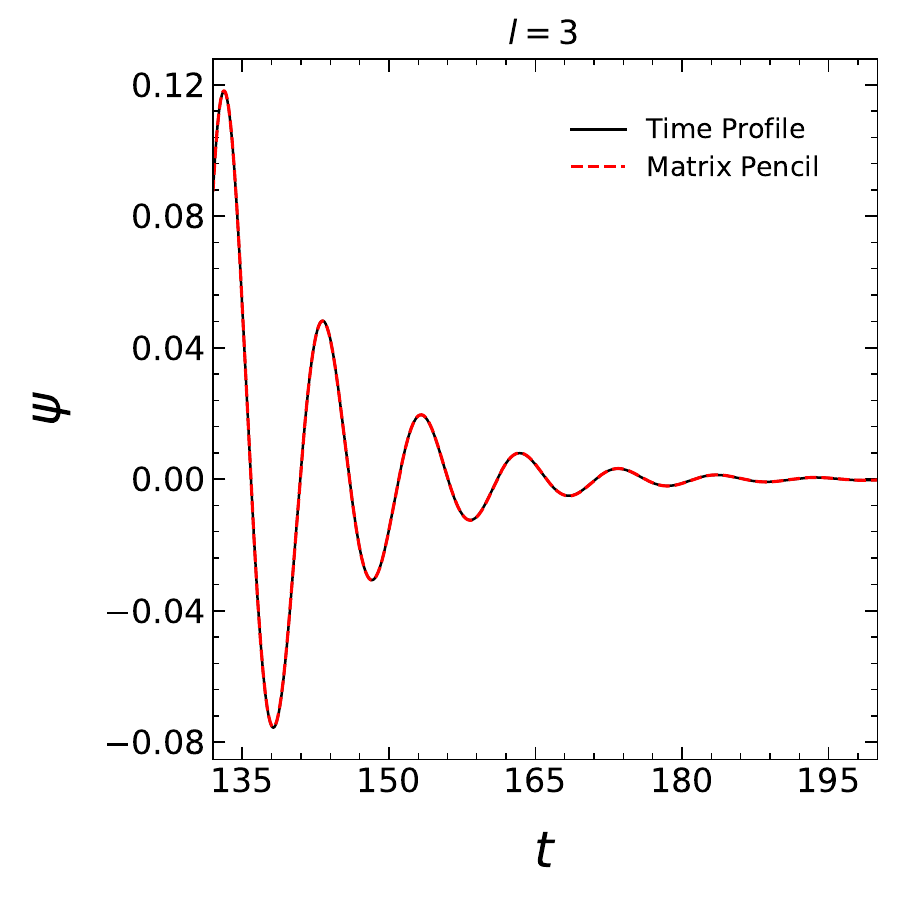}
    \hspace{0.0cm}
    \includegraphics[scale=0.38]{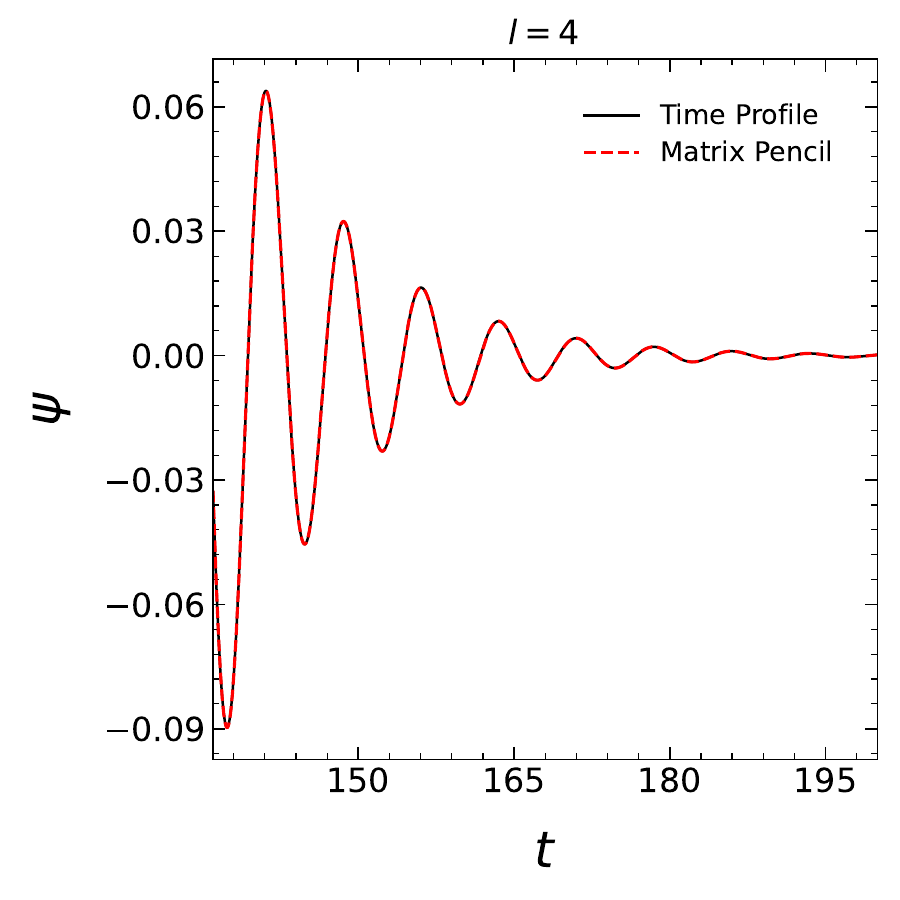}}
    \vspace{-0.2cm}
    \caption{Comparison between the waveforms of the time-domain profile and 
that reconstructed using the Matrix Pencil method with $n=0$, $\xi=0.1$, 
$q=0.5$, $p=0.8$ for different values of $l$ for gravitational perturbations.}
    \label{fig10}
\end{figure}
\begin{table}[htbp]
\centering
\caption{Comparison of the QNM frequencies obtained from the time-domain 
method and the Pad\'e averaged WKB method for different values of $\xi$ and 
$l$ in the case of scalar perturbations. Here, we consider $q=0.5$ and $p=0.8$.}
\label{table3}
\vspace{8pt}
\setlength{\tabcolsep}{12pt}
\begin{tabular}{c c c c c c}
\toprule
$l$ & $\xi$ & Time-domain & Pad\'e averaged WKB & $R^2$ & $\Delta_{\mathrm{QNM}}$ \\
\midrule
\multirow{3}{*}{$l=1$}
&0.1 & $0.302188 - 0.093662i$ & $0.302411 - 0.0940242i$ & $0.999996$ & $0.000212672$\\
&0.3 & $0.298660 - 0.085636i$ & $0.298600 - 0.0862709i$ & $0.999998$ & $0.000318864$ \\
&0.5 & $0.295772 - 0.079825i$ & $0.295819 - 0.0801591i$ & $0.999998$ & $0.000168695$\\
\addlinespace

\multirow{3}{*}{$l=2$}
&0.1 & $0.501109 - 0.092685i$ & $0.501177 - 0.0931996i$ & $0.999998$ & $0.000259537$\\
&0.3 & $0.497800 - 0.085093i$ & $0.498007 - 0.0856127i$ & $0.999996$ & $0.000279704$\\
&0.5 & $0.495612 - 0.079378i$ & $0.495713 - 0.0796196i$ & $0.999999$ & $0.000130931$\\
\addlinespace

\multirow{3}{*}{$l=3$}
& 0.1 & $0.700472 - 0.092552i$ & $0.700578 - 0.0929785i$ & $0.999998$ & $0.000219737$\\
& 0.3 & $0.697283 - 0.085175i$ & $0.697362 - 0.0854389i$ & $0.999999$ & $0.000137735$ \\
& 0.5 & $0.694935 - 0.079125i$ & $0.695045 - 0.0794783i$ & $0.999998$ & $0.000185014$\\

\bottomrule
\end{tabular}
\end{table}
\begin{table}[htbp]
\centering
\caption{Comparison of the QNM frequencies obtained from the time-domain 
method and the Pad\'e averaged WKB method for different values of $\xi$ and 
$l$ in the case of gravitational perturbations. Here we consider $q=0.5$ and 
$p=0.8$.}
\vspace{8pt}
\label{table4}
\setlength{\tabcolsep}{12pt}
\begin{tabular}{c c c c c c}
\toprule
$l$ & $\xi$ & Time-domain & Pad\'e averaged WKB & $R^2$ & $\Delta_{\mathrm{QNM}}$ \\
\midrule
\multirow{3}{*}{$l=2$}
&0.1 & $0.389366 - 0.085191i$ & $0.389264 - 0.0862333i$ & $0.999991$ & $0.00052364$\\
&0.3 & $0.389045 - 0.079546i$ & $0.389204 - 0.0799750i$ & $0.999997$ & $0.00022876$\\
&0.5 & $0.389204 - 0.074396i$ & $0.389294 - 0.0749379i$ & $0.999997$ & $0.00027466$\\
\addlinespace

\multirow{3}{*}{$l=3$}
&0.1 & $0.623071 - 0.089060i$ & $0.623696 - 0.0897004i$ & $0.999994$ & $0.00044742$\\
&0.3 & $0.622690 - 0.082612i$ & $0.623043 - 0.0828466i$ & $0.999998$ & $0.00021192$\\
&0.5 & $0.622246 - 0.077077i$ & $0.622637 - 0.0773626i$ & $0.999996$ & $0.00024210$\\
\addlinespace

\multirow{3}{*}{$l=4$}
& 0.1 & $0.840778 - 0.090922i$ & $0.841309 - 0.0909804i$ & $0.999997$ & $0.00026710$\\
& 0.3 & $0.839579 - 0.083773i$ & $0.839887 - 0.0838695i$ & $0.999999$ & $0.00016138$\\
& 0.5 & $0.838522 - 0.078015i$ & $0.838908 - 0.0782037i$ & $0.999999$ & $0.00015944$\\

\bottomrule
\end{tabular}
\end{table}

In Tables \ref{table3} and \ref{table4}, we show the comparison between the 
QNM frequencies extracted from the time-domain profile using the MP method and 
those obtained from the Pad\'e averaged 6th-order WKB approximation for scalar 
and gravitational perturbations, respectively. From these tables it is seen 
that the QNMs obtained from time-domain analysis show strong consistency with 
those obtained using the Pad\'e averaged WKB approximation. To have a better 
understanding of this consistency, we also calculate the difference between 
the QNMs obtained from time-domain analysis and those obtained from the Pad\'e 
averaged WKB approximation using the formula,
\begin{equation}\label{discrepancy}
\Delta_\text{QNM}=\frac{\left|\mathrm{QNM}_{\mathrm{WKB}}-\mathrm{QNM}_{\mathrm{Time-Profile}}\right|}{2}.
\end{equation}
It is clear that in all the cases values of $\Delta_\text{QNM}$ remain below 
the order of $10^{-3}$.
\section{Possible Black Hole Mass Range for Ringdown Gravitational Wave Detection}\label{sec5}
Here, we discuss briefly about the detection possibilities of QNMs from the 
similar BHs considered in this study. We assume BH mass $M_\text{BH}=
{\eta}M_{\odot}$, where $\eta$ is some poisitive number and 
$M_{\odot}=1.48 \times 10^5$ cm. In physical units, the oscillation frequency 
and decay time of the ringing GWs can be expressed as \cite{mass_range}
\begin{equation}\label{33}
\nu=\frac{cM_\text{BH}\omega_R}{2\pi\eta M_{\odot}}\,\text{kHz}=\frac{32.26 M_\text{BH}\omega_{R}}{\eta}\,\text{kHz},
\end{equation}
and
\begin{equation}\label{34}
\tau=\frac{\eta M_{\odot}}{M_\text{BH}\omega_Ic}=\frac{\eta\times
0.4937\times 10^{-5}}{M_\text{BH}\omega_I}c^{-1},
\end{equation}
where $\omega_R$ and $\omega_I$ are the real and imaginary parts of the QNM 
frequency $\omega$.
Using these expressions, we can check the possibility of detection of GW 
signals from the perturbed BHs, provided we know the sensitive frequency range 
of the detectors. For LISA, the sensitivity range is $\nu \in [10^{-4},\,1]$ 
Hz \cite{mass_range2}. Using this sensitivity range for QNMs shown in Tables 
\ref{table1} and \ref{table2}, we obtain the detectable mass range for the 
scalar perturbations with $l=2$ as
\begin{equation}
1.62 \times 10^{4}\,M_{\odot} \leq M_\text{BH} \leq 1.62 \times 10^{8}\,M_{\odot},
\end{equation}
and for the gravitational perturbations, it is found to be
\begin{equation}
1.25 \times 10^{4}\,M_{\odot} \leq M_\text{BH} \leq 1.25 \times 10^{8}\,M_{\odot}.
\end{equation}
\section{Parameter Estimation from Ringdown Signals}\label{sec6}
In this section, we discuss about the estimation of BH parameters for a given 
detected QNM. To estimate the BH parameters from the ringdown waveform, we 
employ the Fisher analysis technique, which is a widely used technique for 
parameter estimation with future GW detectors \cite{Fisher1, Fisher2, 
ringing_GW}. As mentioned earlier, the ringdown signal is characterized by the 
QNM frequency, which depends on the properties of the underlying 
BH spacetime. \\
\indent If $\tilde{h}(\nu, \theta_i)$ is a ringdown GW form, where $\nu$ is 
the frequency and $\theta_i$ are the parameters on which the GW depends, 
then the root mean squared error on a parameter is
\begin{equation}
\label{eq31}
{\Delta}\theta^i=\sqrt{\Gamma_{ii}^{-1}},
\end{equation}
where $\Gamma_{ij}$ is the Fisher matrix as given by
\begin{equation}
\label{eq32}
\Gamma_{ij} = \left( \frac{\partial \tilde{h}}{\partial \theta_i} \,\middle|\, \frac{\partial \tilde{h}}{\partial \theta_j} \right).
\end{equation}
In the above expression, the notation $(a|b)$ denotes the inner product of
two vectors $a$ and $b$, defined using the detector's power spectral density 
$S_n(\nu)$ as
\begin{equation}\label{33}
(a|b) := 2 \int_{0}^{\infty} 
\frac{a^*b + b^*a}{S_n(\nu)}\,d\nu.
\end{equation}
From the definition of inner product, the signal-to-noise ratio (SNR) of the 
detector can be defined as
\begin{equation}\label{34}
\rho^2 := (\tilde{h}|\tilde{h}) = 4 \,\mathrm{Re}\! \int_{0}^{\infty} \frac{|\tilde{h}(\nu)|^2}{S_n(\nu)}\,d\nu.
\end{equation}
We consider the waveforms to be of the form \cite{ringing_GW}:
\begin{align}
h_+(t)&=\frac{M_z}{D_L}A^+_{lmn}e^{-t/\tau_{lmn}}
\cos\left(2\pi \nu_{lmn} t+\phi^+_{lmn}\right){}_{-2}Y_{\ell m}(\iota,\varphi),
\\[5pt]
h_\times(t)& =\frac{M_z}{D_L}A^\times_{lmn} e^{-t/\tau_{lmn}}
\sin\left(2\pi \nu_{lmn} t+\phi^\times_{lmn}\right){}_{-2}Y_{\ell m}(\iota,\varphi),
\end{align}
where $A^{+,\times}$ and $\phi^{+,\times}$ are the amplitudes and phases of 
the plus and cross GW polarizations, respectively, and 
${}_{-2}Y_{\ell m}(\iota,\varphi)$ is the spin-weighted spherical harmonics of 
spin weight $-2$, which is assumed to be real. $D_L$ is the luminosity
distance, $\nu_{lmn}$ and 
$\tau_{lmn}$ are the detector frame oscillation frequency and damping time of 
the ringing GW strain evaluated using Eqs.~\eqref{33} and \eqref{34}, 
respectively. For fixed values of magnetic charge $p$ and electric charge $q$, 
the frequency and damping time are functions of mass $M_z$ and Lorentz 
violating parameter $\xi$. Here, $M_z$ is the detector frame mass, also known
as redshift mass, and is defined as 
$M_z=(1+z)M_\text{BH}$, where $z$ is the redshift. For simplicity, we consider 
that both the polarizations 
have the same amplitude and phase, i.e., $A^+_{lmn}=A^{\times}_{lmn}=A$ and 
$\phi^+_{lmn}=\phi^{\times}_{lmn}=\phi$. Then the waveform depends on four 
parameters ($A, \phi, M_z, \xi$).\\
\indent The quality factor of a QNM is defined as
\begin{equation}
Q_{lmn} \equiv \pi \nu_{lmn}\tau_{lmn}
= \frac{\omega_{R}}{2\omega_I}.
\end{equation}
Then the amplitude of the GW signal can be defined as \cite{ringing_GW}
\begin{equation}\label{44}
A=\frac{\sqrt{3}}{2}\sqrt{\frac{32\,Q_{lmn}\epsilon_\text{rd}}{M\nu_{lmn}(1+4\,Q_{lmn}^2)}},
\end{equation}
where $\epsilon_\text{rd}$ is the radiation efficiency of the detector. The 
factor $\sqrt{3}/2$ is due to the fact that the LISA arms form an angle of 
$60^\circ$. Eq.~\eqref{44} is used to determine the fiducial amplitude of 
the ringdown waveform for a given choice of $\epsilon_\text{rd}$. In the 
Fisher analysis, however, we will treat $A$ as an independent parameter.\\
\indent For the calculation of SNR and Fisher matrix, it is convenient to 
work in the frequency domain. To calculate the SNR, we follow the prescription 
used by Echeverria \cite{FH_prescription1} and Finn \cite{FH_prescription2} 
and extend the ringdown waveform to $t<0$ by replacing the damping 
factor $e^{-t/\tau_{lmn}}$ with $e^{-|t|/\tau_{lmn}}$. Since the prescription 
effectively doubles the contribution of the signal power, the waveform 
amplitude is divided by a factor of $\sqrt{2}$ to compensate for
doubling the amplitude. The Fourier transformation of the waveform can 
then be obtained using the expression,
\begin{equation}
\int_{-\infty}^{\infty}e^{i2\pi \nu t}\left(e^{\pm i2\pi \nu_{lmn}t-|t|/\tau_{lmn}}\right) dt
=\frac{2/\tau_{lmn}}{\left(1/\tau_{lmn}\right)^2+4\pi^2\left(\nu\pm\nu_{lmn}\right)^2}
\equiv 2b_{\pm}(\nu),
\end{equation}
where
\begin{equation}
b_{\pm}(\nu)=\frac{1/\tau_{lmn}}{\left(1/\tau_{lmn}\right)^2+4\pi^2\left(\nu\pm \nu_{lmn}\right)^2}.
\end{equation}
Therefore, the two polarizations of GWs can be written in the frequency 
domain as
\begin{align}
\tilde{h}_+(\nu)&=\frac{M_z}{D_L} A \left[e^{i\phi} b_+(\nu)+e^{-i\phi} b_-(\nu)\right]{}_{-2}Y_{lm},\\[5pt]
\tilde{h}_{\times}(\nu) & =-i\frac{M_z}{D_L} A \left[e^{i\phi} b_+(\nu)-e^{-i\phi} b_-(\nu)\right]{}_{-2}Y_{lm}.
\end{align}
The waveform detected by the detector is then,
\begin{equation}
\tilde{h}=F_+\tilde{h}_++F_{\times}\tilde{h}_{\times},
\end{equation}
where $F_+$ and $F_\times$ are the pattern functions that depend on the 
orientation of the detector and direction of the source. To calculate SNR and 
Fisher matrix, we average over the source location as well as detector and BH 
orientation, adopting the angle averages 
$\left\langle F_+^2\right\rangle=\left\langle F_\times^2 \right\rangle=1/5$, 
$\left\langle F_+ F_\times \right\rangle=0$, and 
$\left\langle \left|{}_{-2}Y_{lm}\right|^2 \right\rangle=1/4\pi$ \cite{ringing_GW}. 
Thus, using these prescriptions, we get the simplified SNR expression 
as,
\begin{equation}\label{SNR}
\rho^2 =\left(\frac{M_z}{D_L}\right)^2\frac{A^2}{5\pi}\int_{0}^{\infty}
\frac{b_+^2+b_-^2}{S_h(\nu)}\,d\nu,
\end{equation}
\subsection{LISA's Detectability and Signal-to-Noise Ratio}
\begin{figure}[!h]
    \centering
    \includegraphics[scale=0.5]{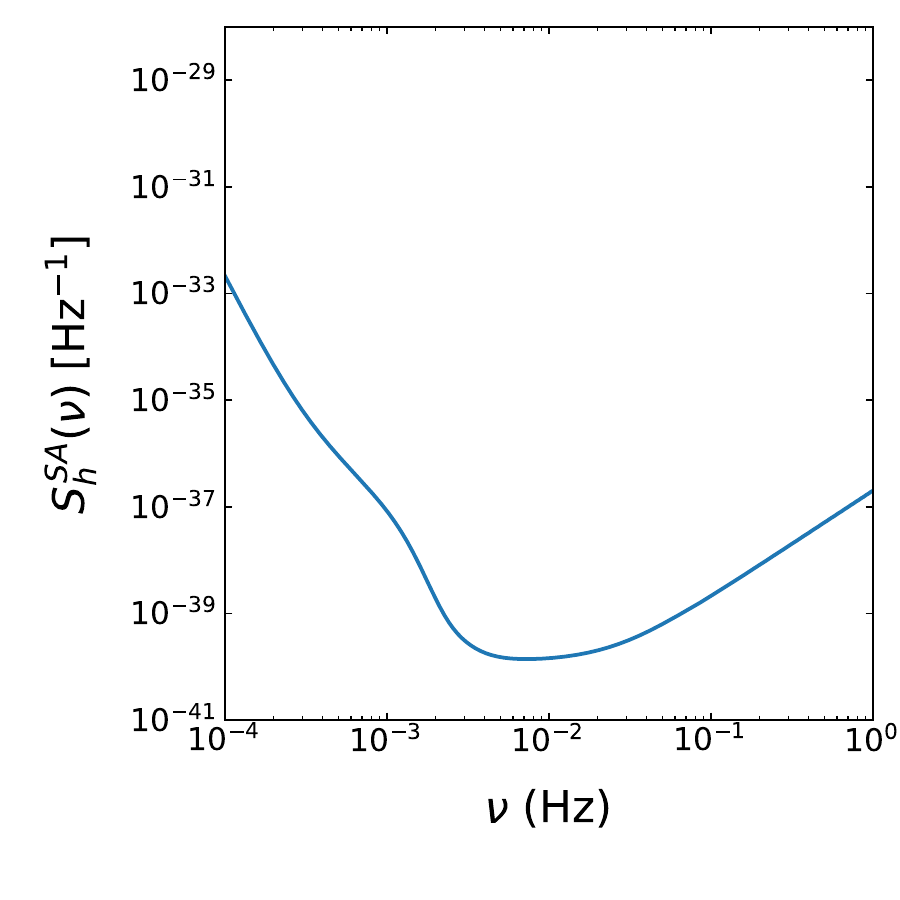}
    \vspace{-0.5cm}
    \caption{Power spectral density (PSD) of LISA as a function of frequency, with galactic confusion noise parameters evaluated for a four-year 
observation period defined in Ref.~\cite{LISA_sensitivity}.}
    \label{fig11}
\end{figure}

To investigate the detectability of the ringdown GW signal, we consider the 
space based GW detector, LISA. By placing three satellites in solar orbit, 
LISA aims to be the first GW observatory in space. It is designed to probe low 
frequency GW signals in the frequency range around $10^{-4}$ Hz to $1$ Hz 
\cite{mass_range2}, much smaller than the typical frequency interval of 
ground-based detectors. According to the latest proposal, presented in 
Ref.~\cite{LISA1}, the observatory is aimed to be launched as early as 2035, 
with a proposed lifetime of 4 years and a possible extension up to 10 years. 
The analytical approximation of LISA's noise curve shown in Fig.~\ref{fig11} 
is defined in Ref.~\cite{LISA_sensitivity} as
\begin{equation}\label{SA_PSD}
S_h^\text{SA}(\nu) = \frac{10}{3L^2}\left(P_{\mathrm{OMS}} + 2\left(1+\cos^2\left(\frac{\nu}{\nu_*}\right)\right)\frac{P_{\mathrm{acc}}}{(2\pi \nu)^4}
\right)\left(1+\frac{6}{10}\left(\frac{\nu}{\nu_*}\right)^2\right) + S_\mathrm{c},
\end{equation}
with optical metrology noise,
\begin{equation}
P_{\mathrm{OMS}} =
\left(1.5 \times 10^{-11}\,\mathrm{m}\right)^2\left(
1+\left(\frac{2\,\mathrm{mHz}}{\nu}\right)^4\right)\,\mathrm{Hz}^{-1},
\end{equation}
and single test mass acceleration,
\begin{equation}
P_{\mathrm{acc}} =
\left(3 \times 10^{-15}\,\mathrm{m\,s^{-2}}\right)^2
\left(1+\left(\frac{0.4\,\mathrm{mHz}}{\nu}\right)^2\right)
\left(1+\left(\frac{\nu}{8\,\mathrm{mHz}}\right)^4\right)
\,\mathrm{Hz}^{-1}.
\end{equation}
Here, $L$ is the arm length and $\nu_* = c/2\pi\nu$ is the transfer frequency. 
For the current LISA design, $L=2.5$ Gm and $\nu_*=19.09$ mHz. Also 
$S_\mathrm{c}$ is the galactic confusion noise which is defined in 
Ref.~\cite{LISA_sensitivity}\\
\indent Now, we investigate the detectability of the ringdown signal by 
calculating the SNR using Eq.~\eqref{SNR}. The power spectral density in 
Eq.~\eqref{SA_PSD} corresponds to sky averaged LISA sensitivity. However, we 
introduced the sky averaging into the definition of the GW signal through the 
detector response. Therefore, we use the corresponding effective 
non-sky-average (NSA) spectral density in SNR calculation and Fisher analysis. 
Following Ref.~\cite{ringing_GW}, we thus define,
\begin{equation}
S_h^\text{NSA}= \frac{3}{20}S_h^\text{SA}.
\end{equation}
\begin{figure}[!h]
    \centering
    \includegraphics[scale=0.45]{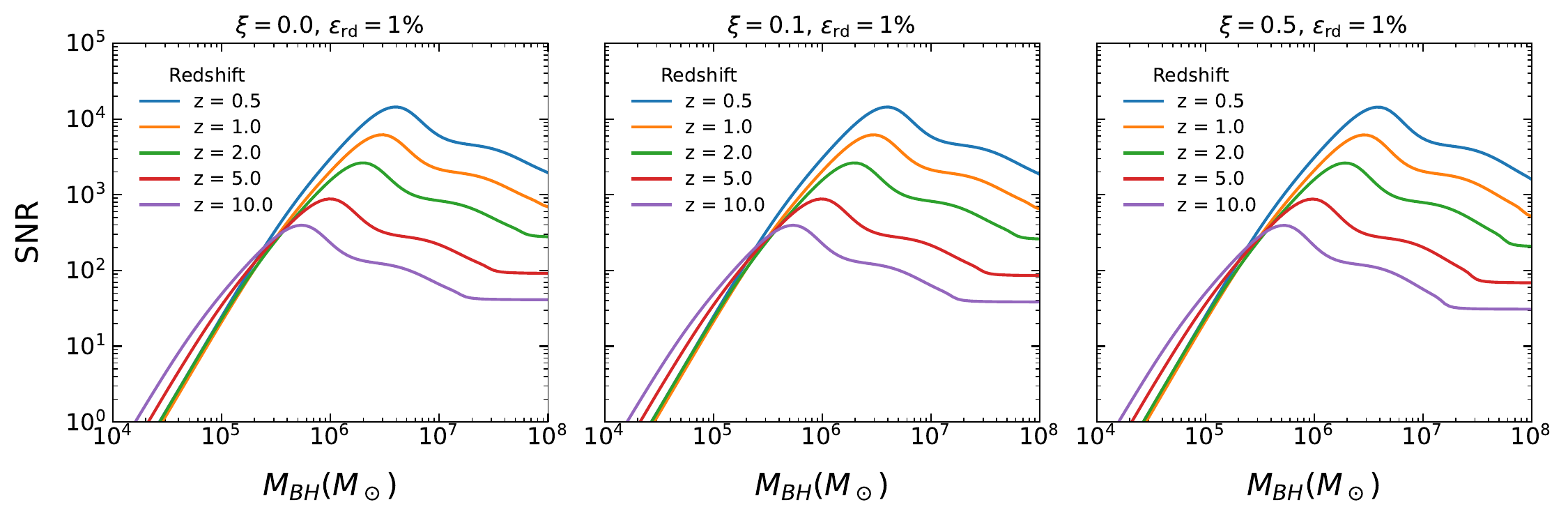}
    \vspace{-0.3cm}
    \caption{Signal-to-noise ratio (SNR) as a function of BH mass for 
ringdown GWs corresponding to scalar perturbations at different redshifts 
with different values of $\xi$. Here, the other parameters are kept fixed at 
$q = 0.5$, $p = 0.8$ and $l = 2$.}
    \label{SNR_scalar}
\end{figure}
\begin{figure}[!h]
    \centering
    \includegraphics[scale=0.45]{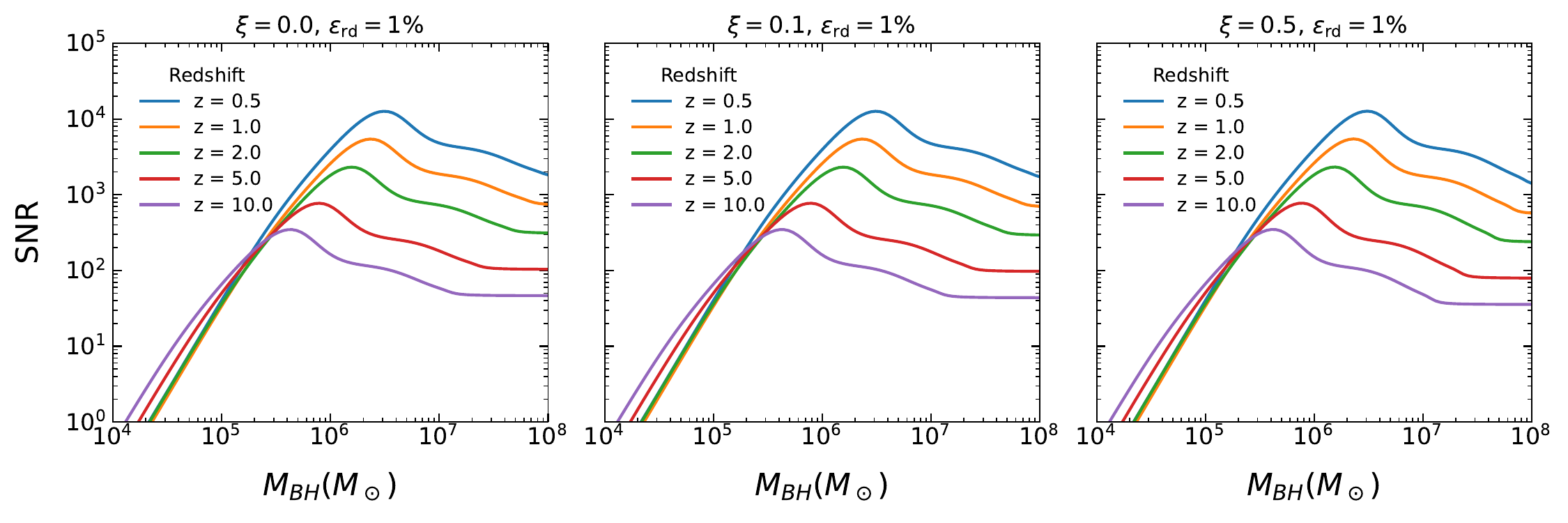}
    \vspace{-0.3cm}
    \caption{Signal-to-noise ratio (SNR) as a function of BH mass for 
ringdown GWs corresponding to gravitational perturbations at different 
redshifts with different values of $\xi$. Here, the other parameters are 
kept fixed at $q = 0.5$, $p = 0.8$ and $l = 2$.}
    \label{fig12}
\end{figure}
\indent We consider the QNMs corresponding to both scalar and gravitational 
perturbations with $l=2$ to examine the detectability of GW signal for 
a range of BH masses corresponding to different redshifts values, viz., 
$z=0.5$, $z=1$, $z=2$, $z=5$, and $z=10$, for three representative values of 
Lorentz violating parameter, $\xi= 0$, $0.1$ and $0.5$, with fixed $q=0.5$ 
and $p=0.8$ values. Throughout the analysis, we choose a ringdown efficiency 
$\epsilon_\text{rd}=0.01$, which fixes the normalization of the ringdown 
amplitude. This value corresponds to a $1\%$ conversion of BH mass into 
ringdown radiation. Fig.~\ref{SNR_scalar} and \ref{fig12} show the resultant SNR as a function of 
BH mass for scalar and gravitational perturbations respectively. For all considered values of $\xi$, the SNR exhibits strong 
dependence on BH mass and redshift. From the figures, it is evident from all 
cases that the LISA is particularly sensitive to ringdown signal coming from 
supermassive BHs, with the largest SNR around 
$M_{\rm BH} \sim 10^6\text{--}10^8\,M_\odot$. Furthermore, the events 
occurring at lower redshifts have high SNR and therefore are more sensitive 
to detection by LISA.
\subsection{Parameter Estimation for LISA}
Following the calculation of SNR, we then investigate the uncertainties in 
the BH parameters $M_\text{BH}$ and $\xi$ using the Fisher analysis 
technique, taking into account the sensitivity power spectral density of 
the LISA detector. In this analysis, we fix the values of electric and 
magnetic charges 
of the BHs as $q=0.5$ and $p=0.8$, which specify a representative dyonic BH 
configuration. Allowing $q$ and $p$ to vary simultaneously would introduce 
additional free parameters in Fisher analysis and is beyond the scope of our 
present analysis. For concreteness, we consider that the source is located at 
$D_L=3$ Gpc corresponding to the redshift $z=0.5$ and has a ringdown 
efficiency $\epsilon_{rd}=1\%$.\\
\begin{figure}[!h]
    \centerline{
    \includegraphics[scale=0.38]{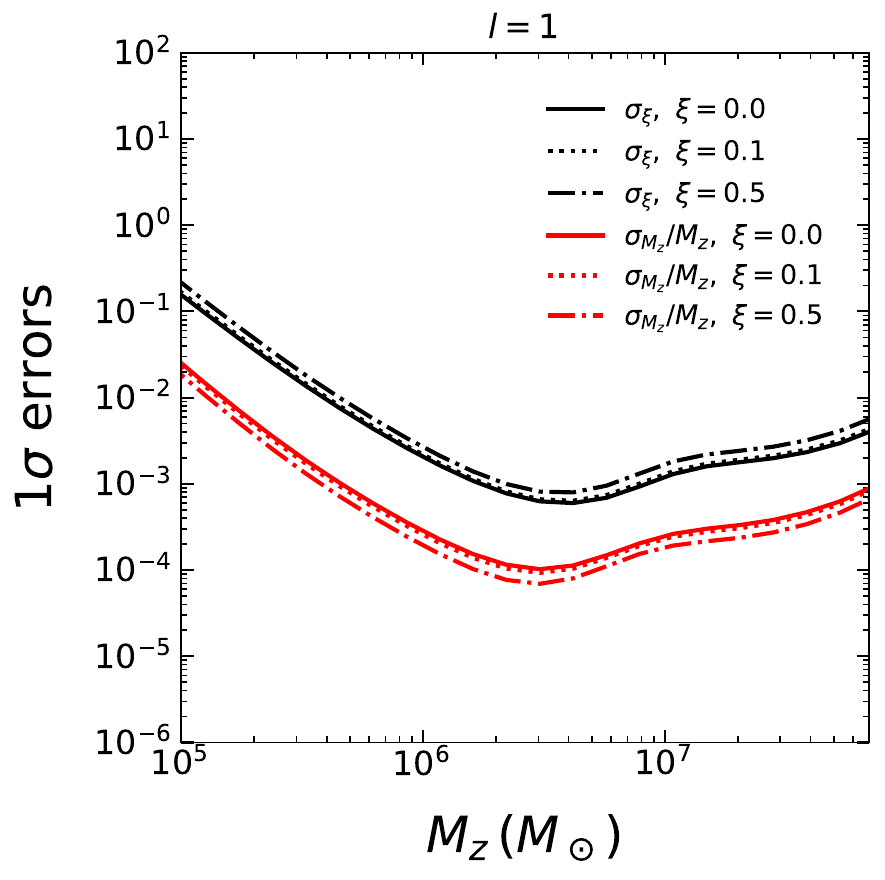}
    \hspace{0.2cm}
    \includegraphics[scale=0.38]{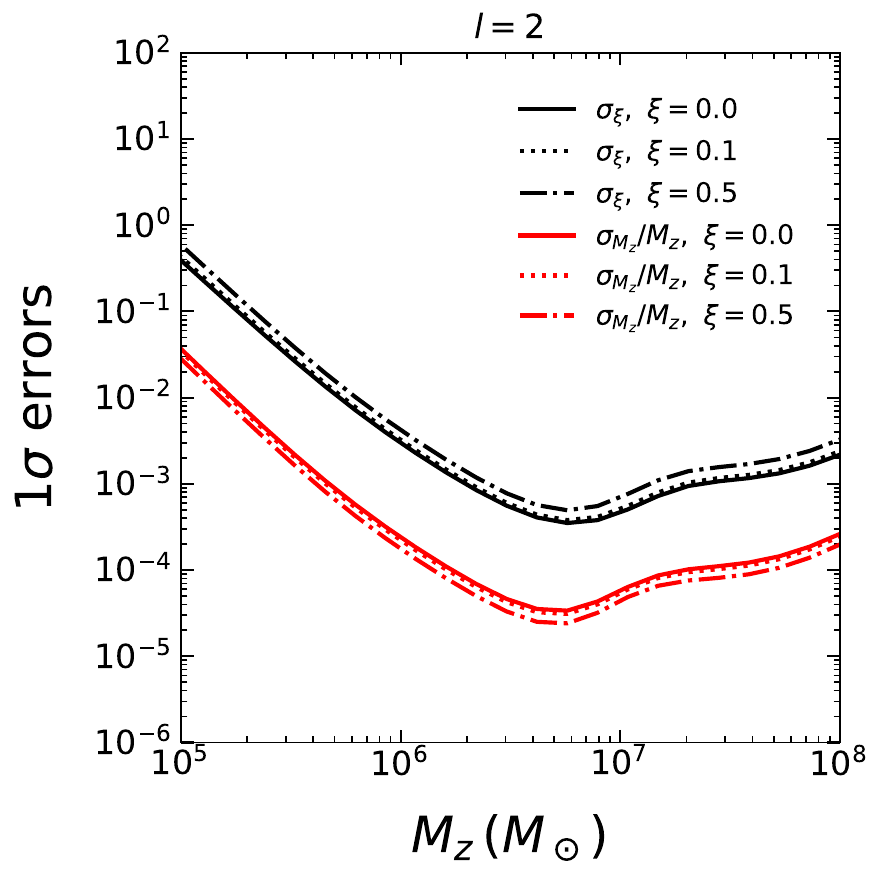}
    \hspace{0.2cm}
    \includegraphics[scale=0.38]{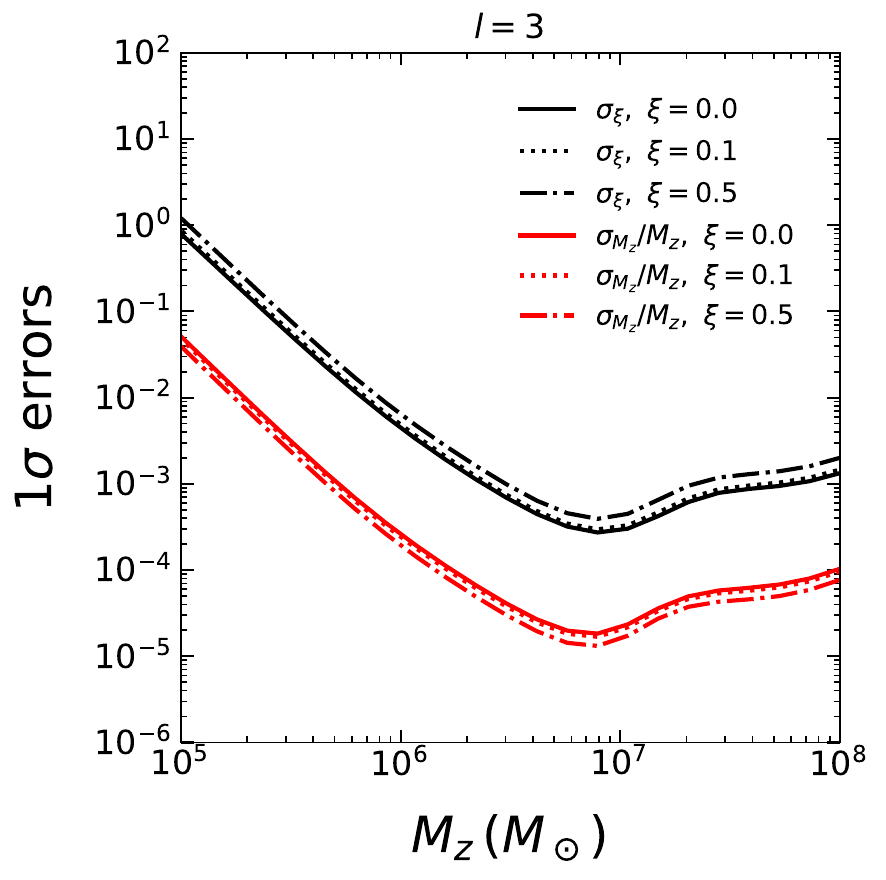}}
    \vspace{-0.2cm}
    \caption{Uncertainty in BH detector frame mass $M_z$ and Lorentz 
violation parameter $\xi$, estimated using QNMs of scalar perturbations for 
a source with $q = 0.5$ and $p = 0.8$ at $D_L=3$ Gpc corresponding to the 
redshift $z=0.5$ and $\epsilon_{rd}=1\%$.}
    \label{fig14}
\end{figure}
\begin{figure}[!h]
    \centerline{
    \includegraphics[scale=0.38]{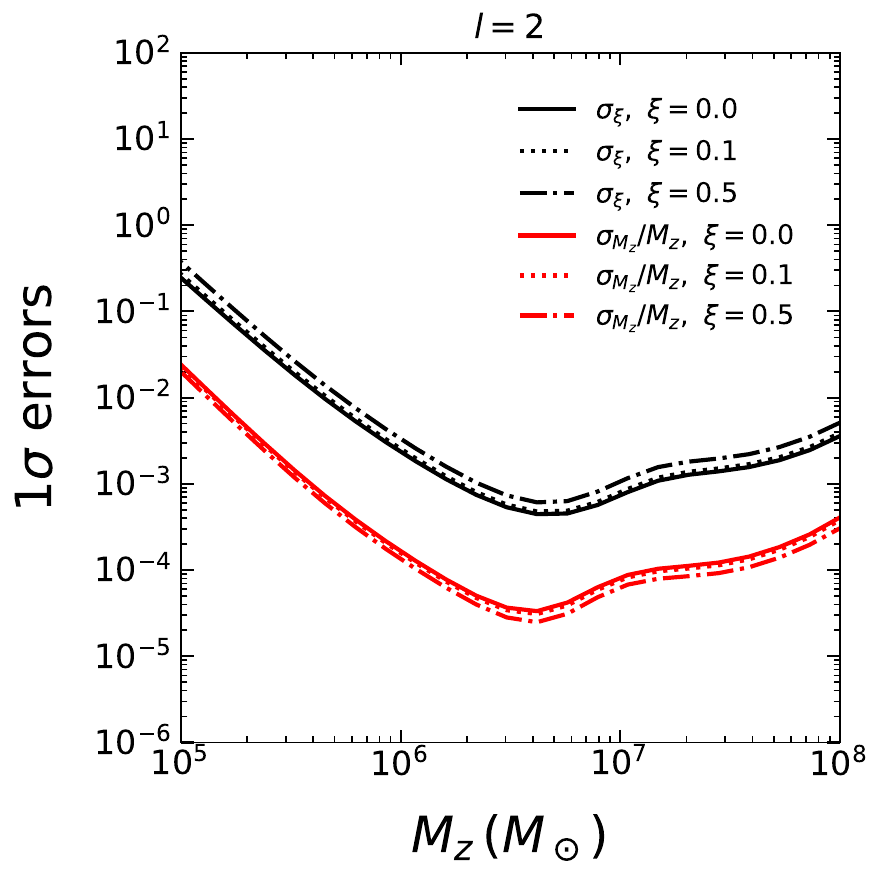}
    \hspace{0.2cm}
    \includegraphics[scale=0.38]{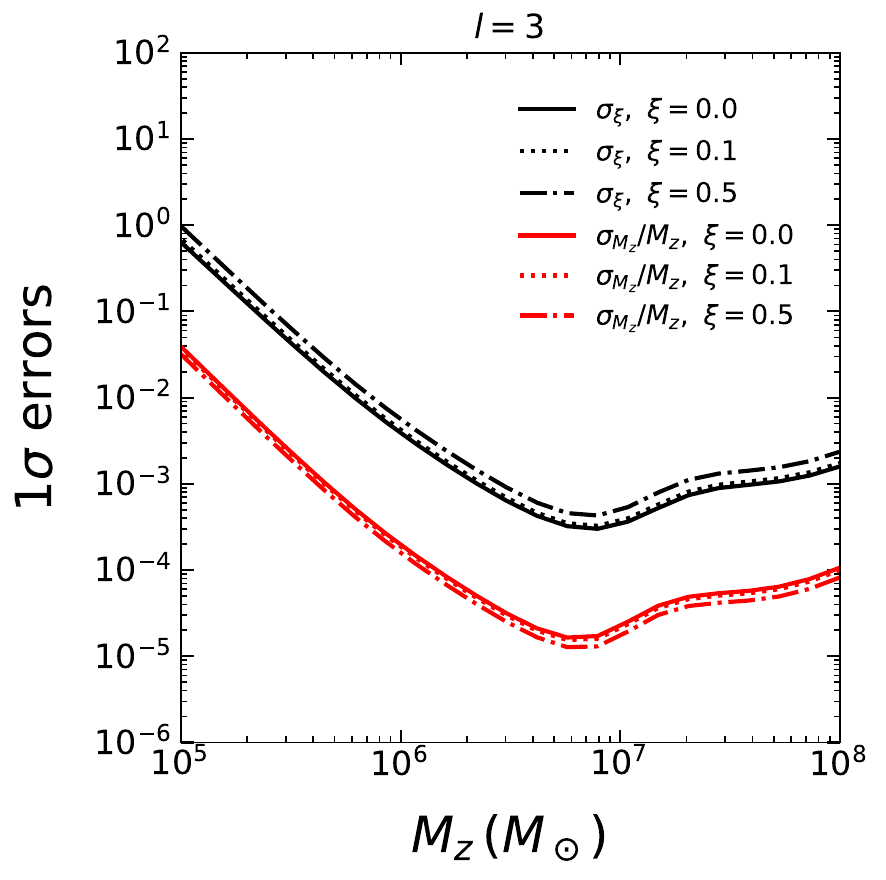}
    \hspace{0.2cm}
    \includegraphics[scale=0.38]{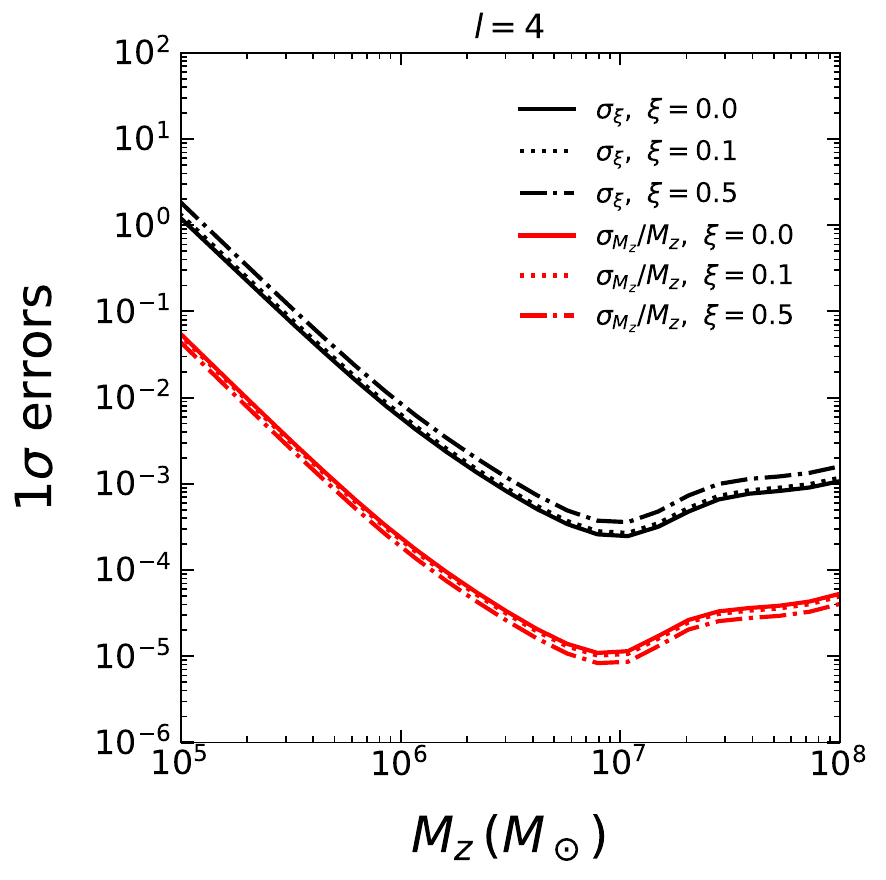}}
    \vspace{-0.2cm}
    \caption{Uncertainty in BH detector frame mass $M_z$ and Lorentz 
violation parameter $\xi$, estimated using QNMs of gravitational 
perturbation for a source with $q = 0.5$ and $p = 0.8$ at $D_L=3$ Gpc 
corresponding to the redshift $z=0.5$ and $\epsilon_{rd}=1\%$.}
    \label{fig15}
\end{figure}
\indent Fig.~\ref{fig14} gives a quantitative idea of the kind of accuracy we 
can achieve if we try to measure mass $M_z$ and Lorentz violating 
parameter $\xi$ for the BHs with LISA. It shows the fractional uncertainty in 
the redshift mass $M_z$ of BHs, $\sigma_M/M_z$ and the uncertainty in 
$\xi$, $\sigma_\xi$, as a function of $M_z$ for three fiducial values 
of $\xi=0$, $0.1$ and $0.5$, considering the multipole indices $l=1$, $2$ and 
$3$ of scalar perturbations. Similarly Fig.~\ref{fig15} shows the uncertainties in parameters for $\xi=0$, $0.1$ and $0.5$, considering the multipole indices $l=2$, $3$ and $4$ of gravitational perturbations. From the figures, we see that the uncertainties in the measurement of the BH parameters from a single 
ringing event are minimum for BHs with mass around 
$10^{6} - 10^{7}M_\odot$ for both cases. Moreover, LISA can constrain the redshift mass $M_z$ 
at the level of $\mathcal{O}(10^{-3})$ or less for BHs with redshift mass 
$M_z\geq 10^6M_{\odot}$. However, the constraint on $\xi$ is 
$\mathcal{O}(10^{-2})$ or smaller for BHs with $M_z\geq 10^6M_{\odot}$. This 
indicates that the LISA's ringdown observations may provide tight constraints 
on the Lorentz violating parameter $\xi$, offering a way to test deviations 
from GR.
\section{Summary and Conclusion}\label{sec7}
In this work, we investigate the QNMs and ringdown GW signals from 
non-rotating dyonic BHs in the framework of EbM gravity theory. We first 
calculate the QNMs of the BHs using the 6th-order Pad\'e averaged WKB 
approximation method by considering scalar and gravitational perturbations and 
study their variation in terms of the BH parameters $\xi$, $q$ and $p$. We 
mainly focus on the effect of Lorentz violating parameter $\xi$ on the QNM 
frequencies. The results show that the QNM spectrum is modified in the presence
of Lorentz violation, indicating that the ringdown signal may carry information 
about the underlying theory of gravity. \\
\indent We then study the time evolution of the perturbed fields around the 
considered BHs and found that the oscillation and decay of QNMs are in good 
agreement with the results obtained from the 6th-order Pad\'e averaged WKB 
approximation method. To extract the QNMs from the time profile, we use the MP 
method. We found that the QNM frequencies obtained in both methods are almost 
equal. The small discrepancy between the two methods is quantified using 
the quantity $\Delta_{\mathrm{QNM}}$ defined in Eq.~\eqref{discrepancy}.\\
\indent We also estimate the BH mass range relevant for ringdown observation 
by comparing the QNM frequencies with the sensitive frequency range of GW 
detectors. We found that QNMs associated with supermassive BHs fall under 
the detectability range of LISA. This motivates us for subsequent parameter 
estimation of the considered BHs using the LISA noise curve.\\
\indent Following this, we constructed  ringdown waveforms corresponding to 
scalar and gravitational perturbations and investigated their detectability by LISA. First 
we calculated the SNR using the LISA noise curve for a range of BH mass and 
redshift, considering different values of the Lorentz violating parameter 
$\xi$. The result shows that the SNR is high for supermassive BHs with mass 
around $10^6 \sim 10^8M_\odot$ and at lower redshifts. Finally, we 
investigate the parameter estimation capability of LISA using the Fisher 
analysis technique. Our result shows that for a representative source at 
$D_L=3$ Gpc 
with $q=0.5$, $p=0.8$ and $\epsilon_{rd}=1\%$, the BH redshift mass $M_z$ and 
the Lorentz violating parameter $\xi$ are well constrained for supermassive 
BHs with $M_{BH}\geq 10^6$. Overall, our results show that the ringdown 
observation of supermassive BHs by LISA may provide a potential probe to test 
the underlying theory of gravity with high precision.\\
\indent In this work, we have restricted our study to only QNM frequencies 
associated with scalar and gravitational perturbations around the considered 
BHs. However, a natural extension of this work would be to study the QNMs 
associated with the electromagnetic perturbation around the BHs. Furthermore, 
in this work we have used only the ringdown GW signals for the parameter 
estimation. A possible future scope for extending the study would be to 
compute bounds on the model parameters using the complete 
inspiral-merger-ringdown signals.

\section*{Acknowledgments}
UDG is thankful to the Inter-University Centre for Astronomy and Astrophysics
(IUCAA), Pune, India, for the Visiting Associateship of the institute.

\end{document}